\documentclass[letterpaper]{jpconf}
\usepackage{graphicx} 
\usepackage{color}
\usepackage{epsfig}

\newcommand{\nlsim}{\mathrel{\rlap{\lower4pt\hbox{\hskip0pt$\sim$}} 
 \raise1pt\hbox{$<$}}}           
\newcommand{\ngsim}{\mathrel{\rlap{\lower4pt\hbox{\hskip0pt$\sim$}} 
 \raise1pt\hbox{$>$}}}           

\def\deg{^\circ}
\def\gtorder{\mathrel{\raise.3ex\hbox{$>$}\mkern-14mu
 \lower0.6ex\hbox{$\sim$}}}
\def\ltorder{\mathrel{\raise.3ex\hbox{$<$}\mkern-14mu
 \lower0.6ex\hbox{$\sim$}}}

\def\mugegm{\mu_p G_E / G_M}
\def\gegm{G_E / G_M}
\def\ge{G_E}
\def\gm{G_M}
\def\mugegmp{\mu_p G_E^p / G_M^p}
\def\gegmp{G_E^p / G_M^p}
\def\gep{G_E^p}
\def\gmp{G_M^p}

\def\gegmn{G_E^n / G_M^n}
\def\gen{G_E^n}
\def\gmn{G_M^n}
\def\q2{$Q^2$}

\def\gev2{GeV$^2$}
\def\3he{$^3$He}

\begin{document}

\title{Nucleon Form Factors - A Jefferson Lab Perspective}

\author{John Arrington$^1$, Kees de Jager$^2$ and Charles F. Perdrisat$^3$}

\address{$^1$ Physics Division, Argonne National Laboratory,
Argonne, IL 60439, USA}

\address{$^2$ Thomas Jefferson National Accelerator Facility, Newport News, VA
23606, USA}

\address{$^3$ College of William and Mary, Williamsburg, VA 23187, USA}

\ead{johna@anl.gov}

\begin{abstract}

The charge and magnetization distributions of the proton and neutron are
encoded in their elastic electromagnetic form factors, which can be measured
in elastic electron--nucleon scattering.  By measuring the form factors, we
probe the spatial distribution of the proton charge and magnetization,
providing the most direct connection to the spatial distribution of quarks
inside the proton.  For decades, the form factors were probed through
measurements of unpolarized elastic electron scattering, but by the 1980s,
progress slowed dramatically due to the intrinsic limitations of the
unpolarized measurements.  Early measurements at several laboratories
demonstrated the feasibility and power of measurements using polarization
degrees of freedom to probe the spatial structure of the nucleon. A program of
polarization measurements at Jefferson Lab led to a renaissance in the field
of study, and significant new insight into the structure of matter.

\end{abstract}

\section{Introduction}\label{sec:introduction}

The electromagnetic form factors encode the spatial distributions of charge
and magnetization in the nucleon.  In a simple picture, the two form factors 
of a spin-1/2 object, $\ge$ and $\gm$, relate to the spatial distribution of
charge and magnetization inside the object.  In the nucleon, the quarks
are the carriers of charge, and so these observables are directly connected
to the spatial distribution of quarks in the nucleon, as well as a probe
of the underlying dynamics.

In the 1950s electron scattering became the tool of choice for measuring the
nucleon form factors.  An active program mapped out the form factors as well
as possible, but the ability to extract the form factors using unpolarized
cross-section measurements was limited.  By the 1980s most of the experiments
provided only incremental improvements on the precision or on the $Q^2$-range of
existing measurements, or were early proof-of-principle tests of new
techniques.  

The advent of electron beams with high luminosity and polarization, combined
with new polarized targets, recoil polarimeters, and large-acceptance
detectors, led to a revolution in the study of nucleon form factors.  In the
last 10 years measurements at Jefferson Lab have rewritten the textbook
on the proton and neutron form factors.  The techniques that have
allowed this dramatic resurgence of interest in the form factors have also
opened up other possibilities, allowing us to try and isolate the contribution
of strangeness in the nucleon and making cleaner and more precise measurements
of the impact of the nuclear environment on the internal structure of the
proton and neutron.

\section{Historical Context}\label{sec:history}

In the Born approximation, where the interaction occurs via the exchange
of a single virtual photon, the unpolarized $e$--$N$ elastic cross
section can be written in terms of the Sachs form factors, $\ge$ and $\gm$, as
\begin{equation}
\frac{d\sigma}{d\Omega} = \frac{\sigma_{Mott}}{\varepsilon (1+\tau)}
\left[ ~ \tau \gm^2(Q^2) + \varepsilon \ge^2(Q^2) ~ \right]~,
\label{eq:rosenbluth}
\end{equation}
where $\tau=Q^2/4M^2$, $-Q^2$ the square of the four-momentum transfer, $M$
the nucleon mass, and $\varepsilon = 1/[1 + 2(1 + \tau)
\tan^2(\frac{\theta_e}{2})]$ is the linear polarization parameter of the
virtual photon.  The value of $\varepsilon$ depends on the scattering angle
$\theta_e$, with $\varepsilon \to 1$ in the limit of forward scattering, and
$\varepsilon=0$ for 180$\deg$ scattering.  The term $\sigma_{Mott}$ denotes
the cross section for the scattering of two spin-1/2 point-like objects.  The
quantity in brackets is referred to as the reduced cross section which, at
fixed $Q^2$, depends only on the values of $\ge$, $\gm$, and $\varepsilon$.  
The electric and magnetic form factors can also be written in terms of the
Dirac and Pauli form factors: $\ge = F_1 - \tau F_2$, $\gm = F_1 + F_2$.
Writing the cross section in terms of the Sachs form factors
(Eq.~\ref{eq:rosenbluth}) yields a simpler expression, without the cross terms
that appear when using $F_1$ and $F_2$.  By performing a Rosenbluth separation
- making cross-section measurements at a fixed $Q^2$-value but two or more
$\varepsilon$-values - one can separate the values of $\ge^2$ and
$\gm^2$.

\begin{figure}[h]
\begin{center}
\epsfig{file=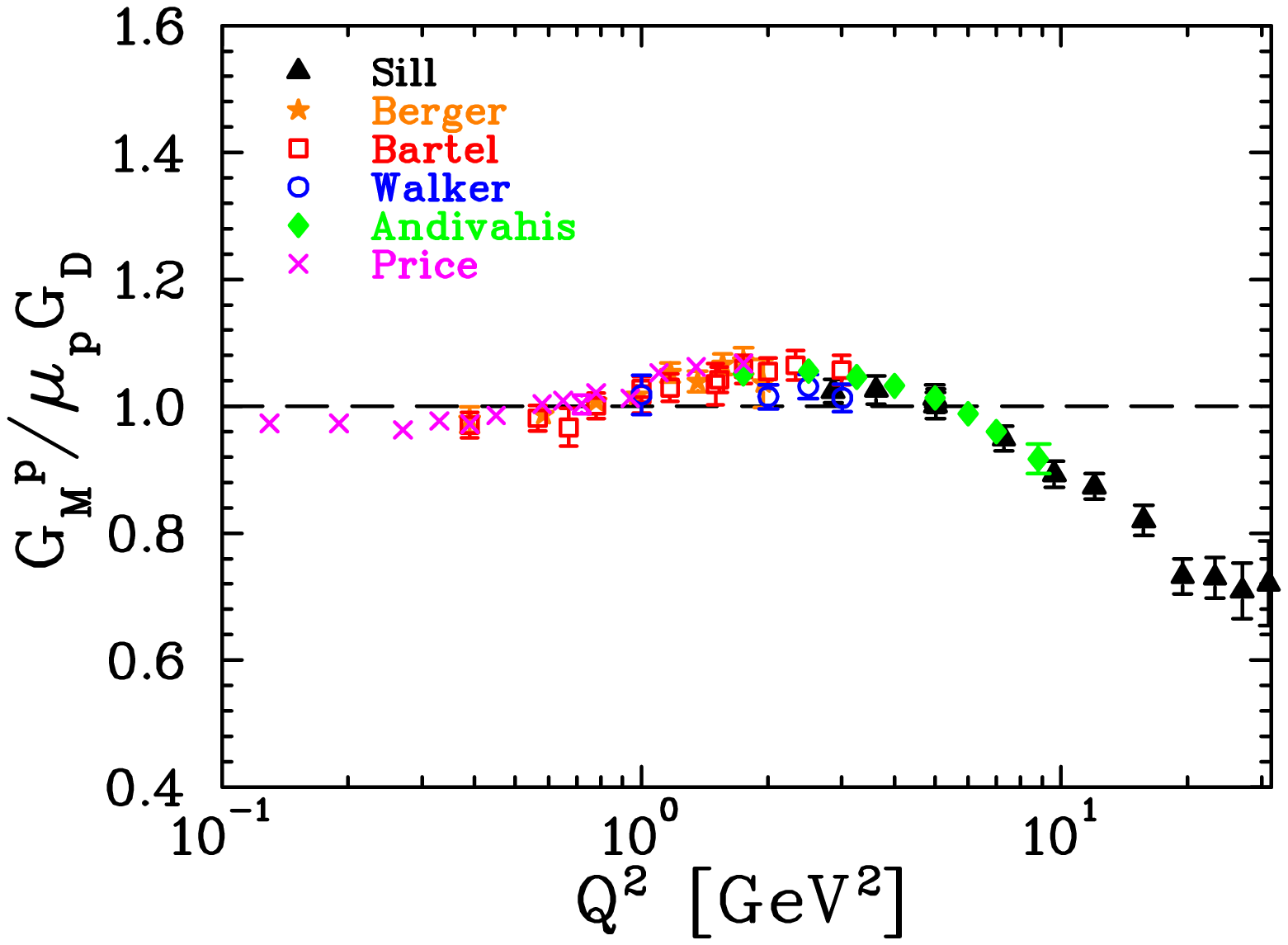,width=0.48\linewidth}
\epsfig{file=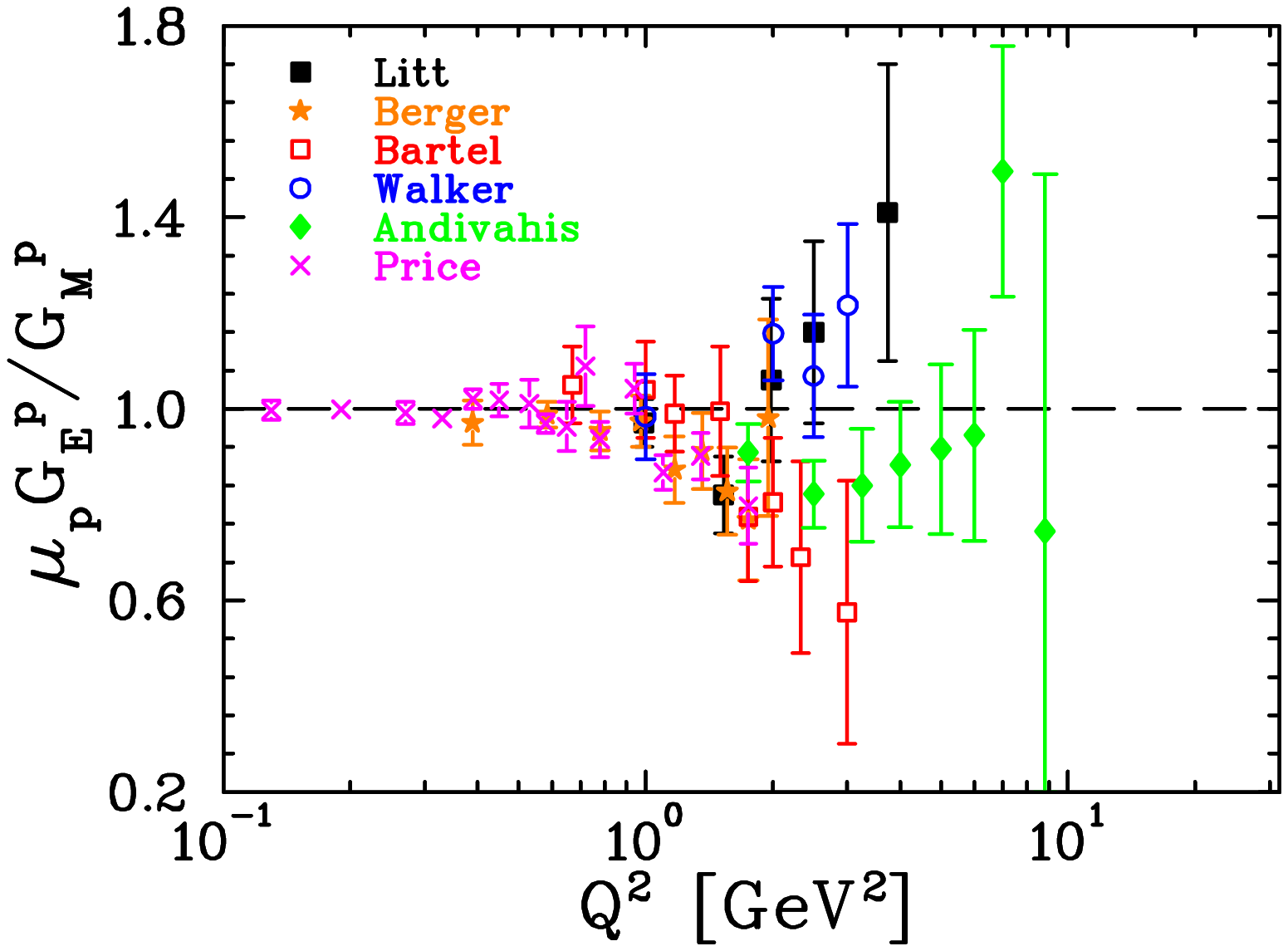,width=0.48\linewidth}
\caption[]{Database for $\gmp/\mu_p G_D$ (left) and $\mugegmp$ (right) 
obtained by the Rosenbluth method as of the mid-1990s.}
\label{fig:proton_early}
\end{center}
\end{figure}

This technique is limited in its ability to make a clean separation of the
form factors.  One can see from the form of the reduced cross section that
there is very little sensitivity to $\ge$ for $Q^2 \gg 1$ -- or for $(\gegm)^2
\ll 1$ -- and little sensitivity to $\gm$ for $Q^2 \ll 1$, except for
$\theta_e \to 180\deg$.  
Nonetheless, it was possible to make measurements of $\gep$ and especially
$\gmp$ over a wide range in $Q^2$, as seen in Fig.~\ref{fig:proton_early},
which shows the status of proton 
form-factor measurements in the mid-1990s, see~\cite{hydewright04, perdrisat07,
arrington07a} for details of the measurements.  The $Q^2$-dependence of the
proton magnetic form factor is well approximated by the dipole form up to
10~GeV$^2$ ($\gmp/\mu_p \approx G_D = (1+Q^2/0.71)^{-2}$ with $\mu_p$ the
proton magnetic moment), and falls $\sim$30\% below the dipole form at $Q^2
\approx 30$~GeV$^2$.   While $\gep$ was also reasonably well approximated by
the dipole form, systematic variations between the results of different
experiments were much larger in $\gep$ than $\gmp$.  However, the general
conclusion was that the data were consistent with form-factor scaling, i.e. 
$\mugegmp$ was independent of $Q^2$, up to at least 5 GeV$^2$.  This was
consistent with simple non-relativistic quark models, as well as the
perturbative QCD expectation at large $Q^2$.

\begin{figure}[h]
\begin{center}
\epsfig{file=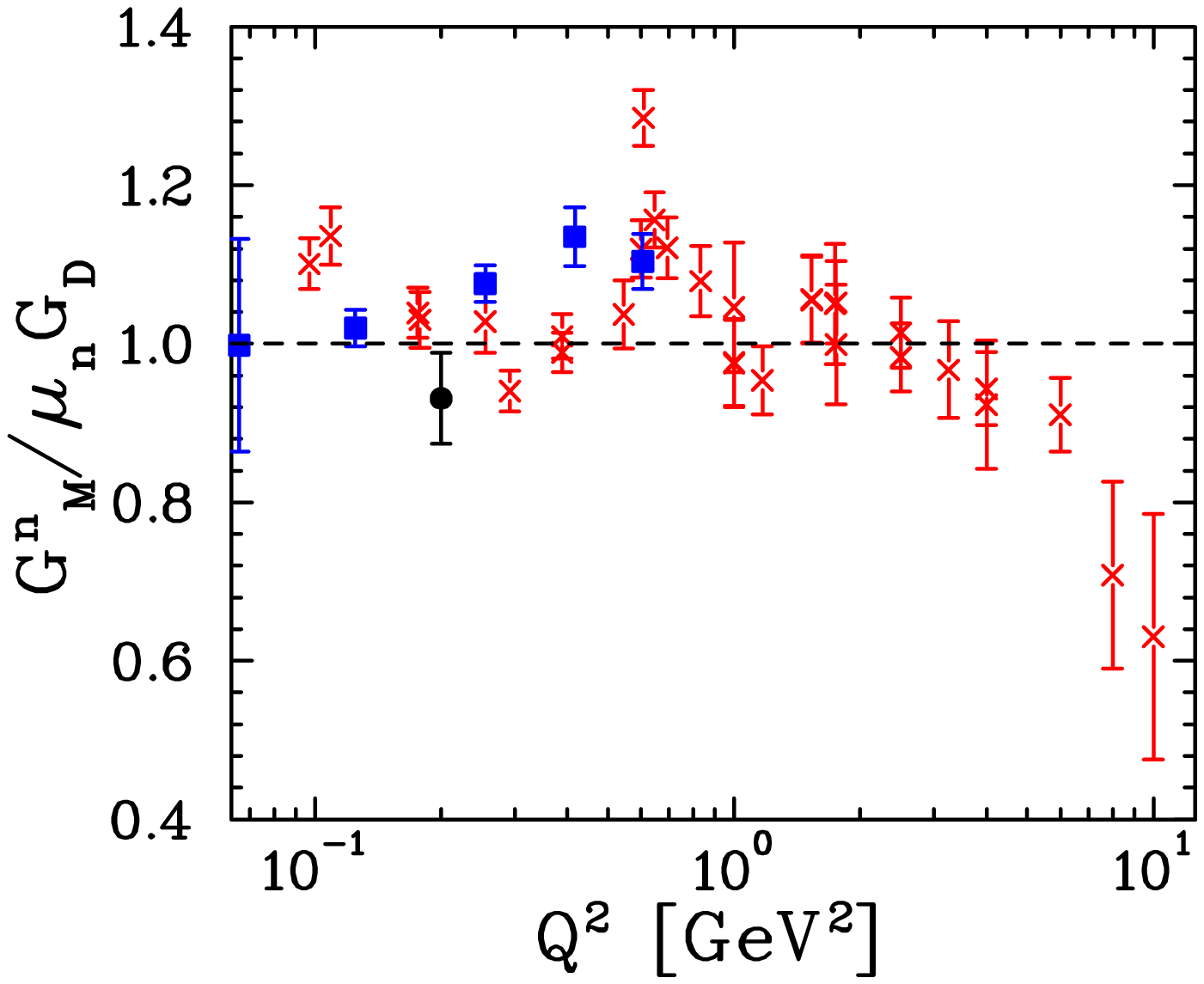,width=0.48\linewidth}
\epsfig{file=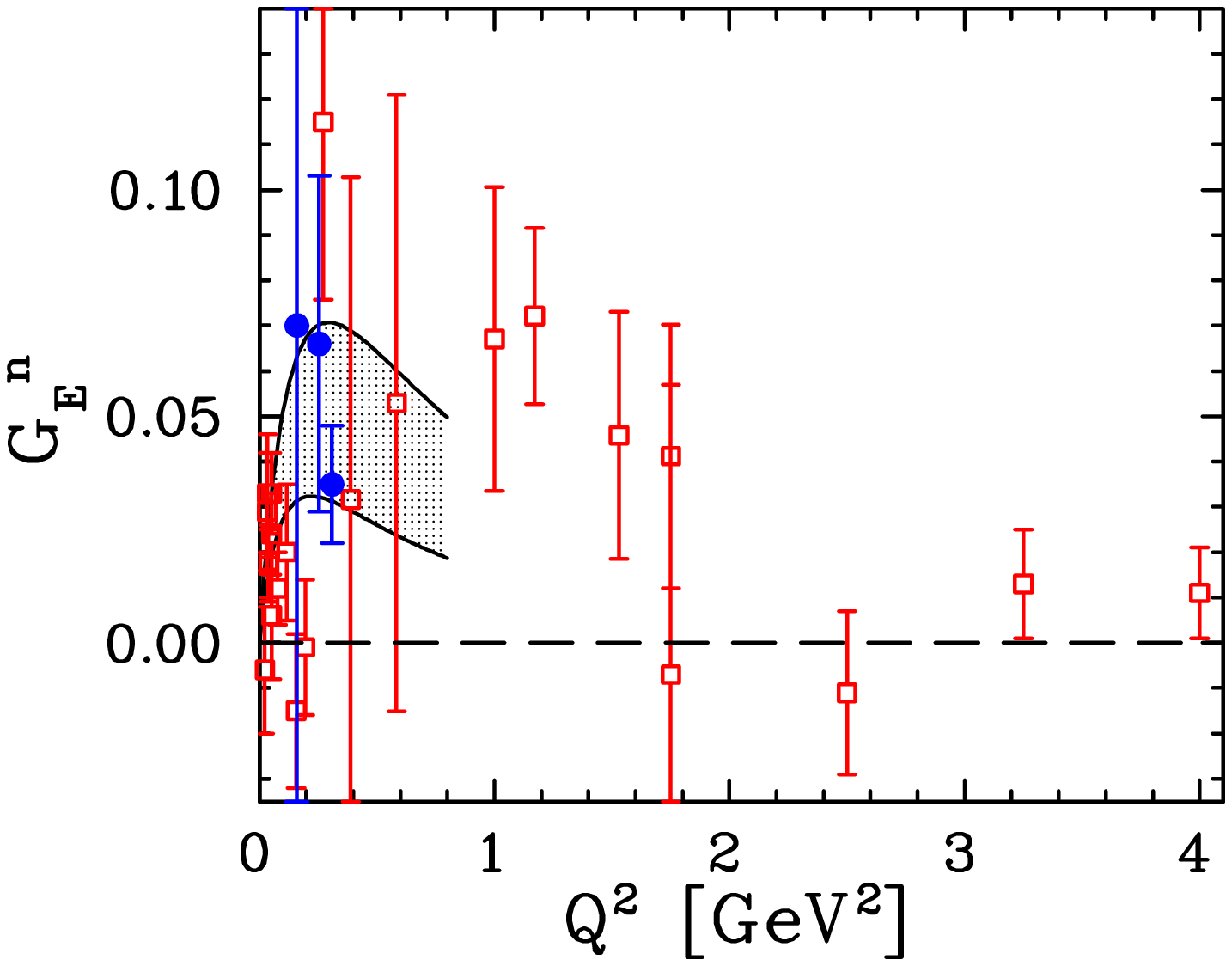,width=0.48\linewidth}
\caption[]{Database for $\gmn/\mu_n G_D$ (left) and $\gen$ (right) before the
JLab turn-on.  For $\gmn$, the crosses are from quasi-elastic scattering
(inclusive and coincidence), the solid squares are from ratio measurements,
and the solid circle from a polarization experiment. For $\gen$, the solid circles are from
polarization measurements, and the hollow squares from quasi-elastic
scattering data. The shaded band shows the range of results extracted from a
model-dependent analysis of elastic $e$--$d$ scattering~\cite{platchkov90}.}
\label{fig:neutron_early}
\end{center}
\end{figure}

For the neutron, measurements using the Rosenbluth technique were even more
difficult.  The need to use light nuclei as ``effective'' neutron targets 
necessitated large corrections in extracting the $e$--$n$ cross section, 
thus limiting the precision and kinematic coverage for extractions of $\gmn$. 
The extraction of $\gen$ was even more problematic, as $\gen$ is much smaller
than the other electromagnetic form factors.  Therefore, there were no precise
extractions of $\gen$, and for $Q^2 \gtorder 2$~GeV$^2$ only upper limits
could be set.  In the limit $Q^2 \to 0$, $\gen$ must approach zero (the charge
of the neutron),  while at finite $Q^2$, any non-zero value must come from a
difference in the spatial distribution of up and down quarks in the neutron.

Figure~\ref{fig:neutron_early} shows the status of neutron form-factor
measurements around the time of the JLab turn-on in 1996. For $\gen$, the data
are mainly from quasi-elastic scattering from the deuteron and from elastic $e$--$d$
scattering, which has large model-dependent uncertainties, indicated by
the shaded band.  Only a few proof-of-principle polarization
measurements had been performed.  For $\gmn$, most of the results are from
inclusive quasi-elastic scattering measurements from the deuteron, and the large
uncertainties show the impact of the proton subtraction and the nuclear
corrections.  A few data had been measured using the so-called ratio technique which,
along with the polarization measurements, will be described in the following
sections.

It had long been known~\cite{akhiezer58, akhiezer68, dombey69, arnold81} that,
in principle, including measurements of polarization 
observables would allow a much improved determination of the form factors. 
Using a polarized electron beam, one can measure either the cross-section
asymmetry from a polarized target or the polarization transferred to an
unpolarized nucleon.  In both cases, the measured asymmetry depends only on
the ratio $\gegm$, thus providing sensitivity to the electric form factor even
if its contribution to the unpolarized cross section is extremely small. 
However, it required the development of high
intensity, highly polarized electron beams and polarized targets or recoil
polarimeters with high figures of merit to apply this theory to advance our
knowledge of the form factors.

\begin{table}
\caption{List of JLab experiments related to nucleon form factors. Experiments
marked with ``$^\dag$'' were focused on two-photon exchange contributions.
Proposals are available from
``http://www.jlab.org/exp\_prog/generated/approved.html''.}
\label{tab:experiments}
\begin{tabular}{|l|l|c||l|l|c|}
\hline
Expt.(Hall)	&	Reaction	& Goal, $Q^2$-range &
Expt.(Hall)	&	Reaction	& Goal, $Q^2$-range \\ \hline
E93-027(A)	& $p(\vec{e},e'\vec{p})$	& $\gep$, 0.5--3.5 &
E95-001(A)	& $\vec{^3\mbox{He}}(\vec{e},e')$	& $\gmn$, 0.1--0.6 \\
E99-007(A)	& $p(\vec{e},e'\vec{p})$	& $\gep$, 3.5--5.6 &
E94-017(B)	& $^2$H$(e,e'N)$		& $\gmn$, 1--4.8 \\
E04-108(C)	& $p(\vec{e},e'\vec{p})$	& $\gep$, 2.5--8.5 &
E93-026(C)	& $\vec{^2\mbox{H}}(\vec{e},e'n)$ & $\gen$, 0.5--1.0 \\
E04-019(C)	& $p(\vec{e},e'\vec{p})$$^\dag$	& $\gep$, 2.5 &
E93-038(C)	& $^2$H$(\vec{e},e'\vec{n})$	& $\gen$, 0.4--1.5 \\
E01-001(A)	& $p(e,e'p)$$^\dag$		& $\gep$, 2.6--4.1 &
E02-013(A)	& $\vec{^3\mbox{He}}(\vec{e},e'n)$	& $\gen$, 1.4--3.4 \\
E05-017(C)	& $p(e,e'p)$$^\dag$		& $\gep$, 0.4--5.8 &
E04-110(C)	& $^2$H$(\vec{e},e'\vec{n})$	& $\gen$, 4.3\\
E08-007(A)	& $p(\vec{e},e'\vec{p})$	& $\gep$, 0.25--0.7 &
E07-005(B)	& $p(e^\pm,e'p)$$^\dag$		& TPE, 0.5--3.0 \\
E08-007(A)	& $\vec{p}(\vec{e},e'p)$	& $\gep$, $$0.015--0.4 &
 &  & \\
\hline
\hline
\end{tabular}
\end{table}

Already at the third Program Advisory Committee (PAC) meeting of the CEBAF in
February of 1989 three letters of intent (LOI) had been presented to measure
nucleon form factors with the new techniques being developed; two to measure
$\gen$ and one to measure $\gep$.  These  LOIs were then developed into
full-blown proposals, resulting in five fully approved proposals in 1994, 
shortly before the start of operations. The first experiments provided much
improved measurements of proton and neutron form factors, as well as some
unexpected and exciting results.  This led to a variety of proposals,
summarized in Table~\ref{tab:experiments}, aimed at fully developing and
exploiting these new techniques to extend the kinematic coverage and precision
of the data.  Within a decade of the first measurement, this program would
drastically transform our state of understanding of nucleon form factors.
Related measurements examining parity-violating elastic scattering as well as
meson and transition form factors are discussed in other reviews in this
collection.

\section{Proton Form Factors}\label{sec:proton}

The original motivation for the new measurements of the proton electric
form factor was the internal inconsistency of the database of the time, 
which showed a rapid increase in the uncertainty and scatter of the results
with $Q^2$, starting at $\sim$1~GeV$^2$; see Fig.~\ref{fig:proton_early}. The
initial experiment in Hall A yielded unexpected results, which created
significant excitement in the field.  A great deal of effort went into 
trying to understand the implications of the new data on our picture of
the proton, and to extend the polarization measurements to higher $Q^2$-values.  
There was also significant activity aimed at understanding the
discrepancy with previous measurements, and making significant improvements
in precision at low $Q^2$-values.

First, we discuss the new techniques that made such measurements feasible, then
we present the results of the initial high-$Q^2$ experiments, as well as the
later studies that were an offshoot of these new techniques.

\subsection{Techniques}\label{sec:ptech}

Polarization-transfer experiments measure the polarization of the recoiling
proton by re-scattering it in an appropriate material and determining the
resulting azimuthal asymmetry distribution, thus providing a measurement of
the two components of the polarization in a plane perpendicular to the proton
momentum.  After being struck by a polarized electron, the
proton has in-plane polarization components
parallel ($P_{\ell}$) and perpendicular ($P_t$) to the proton momentum. The
component normal to the scattering plane is zero in the Born approximation.  To
measure both $P_{\ell}$ and $P_t$ simultaneously requires precession of
$P_{\ell}$ into a normal component, using  a dipole magnet. The precession in
the dispersive plane of a dipole is given by
$\chi=\gamma\cdot(\theta_B+\theta_{target}-\theta_{fpp})\cdot\kappa_p$, where
$\gamma$, $\theta_B$, $\theta_{target}$, $\theta_{fpp}$ and $\kappa_p=\mu_p-1$
are the proton's relativistic factor, the mean bending angle of the dipole, the
entrance and exit angle of an individual proton trajectory, and the anomalous
proton magnetic moment, respectively. The resulting distribution in the
azimuthal angle $\varphi$ is
\begin{equation}
N(\vartheta,\varphi) = N_0 \cdot
[1+h(P_bA_y(\vartheta)P_{n}^{fpp}\sin\varphi-P_bA_y(\vartheta)P_t^{fpp}\cos\varphi)],
\label{aziasy}
\end{equation} 
where $N_0$ is the average number of events in a given interval of $\theta$ and
$\varphi$, $P_b$ is  the electron beam longitudinal polarization and $h=\pm1$ 
the helicity state of the beam, and $P_n^{fpp} \approx P_{\ell}\sin\chi$ and  
$P_t^{fpp}\approx P_t$ are the
polarization components in the polarimeter; $A_y(\vartheta)$ is the analyzing power at a
given polar scattering angle $\vartheta$. The relative beam-helicity
difference distribution in a given $\vartheta$ interval is then:
\begin{equation}
\frac{N^{h=+1}(\varphi)-N^{h=-1}(\varphi)}{2N_0} = P_b A_y (P_{n}^{fpp}\sin\varphi-P_t^{fpp}\cos\varphi)
\label{eq:heldiff}
\end{equation}
Taking only the spin precession in the dispersive plane, $\chi$, into account, the ratio
$P_t^{fpp} \sin\chi/P_n^{fpp}$ is directly related to the ratio $P_t/P_{\ell}$
at the target, which in turn is a measure of $\gegmp$:
\begin{equation}
\frac{\gep}{\gmp}=-\frac{P_t}{P_{\ell}}\frac{(E_e+E_{e'})}{2M}\tan(\theta_e/2)
=-\frac{P_t}{P_{\ell}}\sqrt{\frac{\tau(1+\varepsilon)}{2\varepsilon}}
\label{ratiop}
\end{equation}

As $\gegmp$ is defined by the ratio of two polarization components, knowledge
of the beam polarization and polarimeter analyzing power is not necessary. 
The remaining source of systematic uncertainty comes from the accuracy of the
spin transport, i.e. the calculation of the components of the polarization at
the target from the asymmetry in the focal plane. As $Q^2$ increases, the
precession in the non-dispersive plane due to focusing elements in the
spectrometer becomes significant and has to be taken into account. This
residual systematic uncertainty can be evaluated on the basis of optical
studies~\cite{punjabi05}.

In almost all recoil-polarization experiments at JLab, both final-state
particles were detected to reduce the inelastic contamination. The first
focal-plane polarimeter (FPP) built in Hall A was used for the GEp(I)
experiment, E93-027~\cite{jones00,punjabi05}.  Several changes were required to utilize the full
beam energy of the accelerator and extend the measurements
to higher $Q^2$. For a fixed beam energy, the cross section scales as
$Q^{-12}$ and the analyzing power decreases with increasing $Q^2$.
Furthermore, the electron solid angle matching the proton acceptance increases
with $Q^2$. To compensate for these factors in GEp(II) (E99-007), the electron
was detected in a large solid-angle electromagnetic calorimeter, and the
polarimeter was reconfigured with 95~g/cm$^2$ of CH$_2$ instead of graphite to
increase the effective analyzing power.

As the maximum momentum of the proton spectrometer in Hall A limited the range
of possible $Q^2$, GEp(III) (E04-018) used the HMS spectrometer in Hall C,
which can detect proton momenta up to 7.5~GeV/c. To increase the figure of
merit at high $Q^2$, a new FPP was built, consisting of two polarimeters in
succession, each containing a slab of polyethylene, 50~g/cm$^2$ thick,
followed by a set of drift chambers.  A larger calorimeter (``BigCal'') was needed to
match the proton acceptance at these large $Q^2$-values. An entirely new
calorimeter was built, consisting of 1744 lead glass bars 4$\times$4 cm$^2$ in cross
section, and 40-45 cm long, with a frontal area of 2.6 m$^2$. A trigger signal
from the calorimeter was required for the definition of an $(e,e'p)$ or an
$(e,\gamma p)$ event.

\subsection{High-$Q^2$ regime}\label{sec:presults1}

\begin{figure}[hbt]
\begin{center}
\epsfig{file=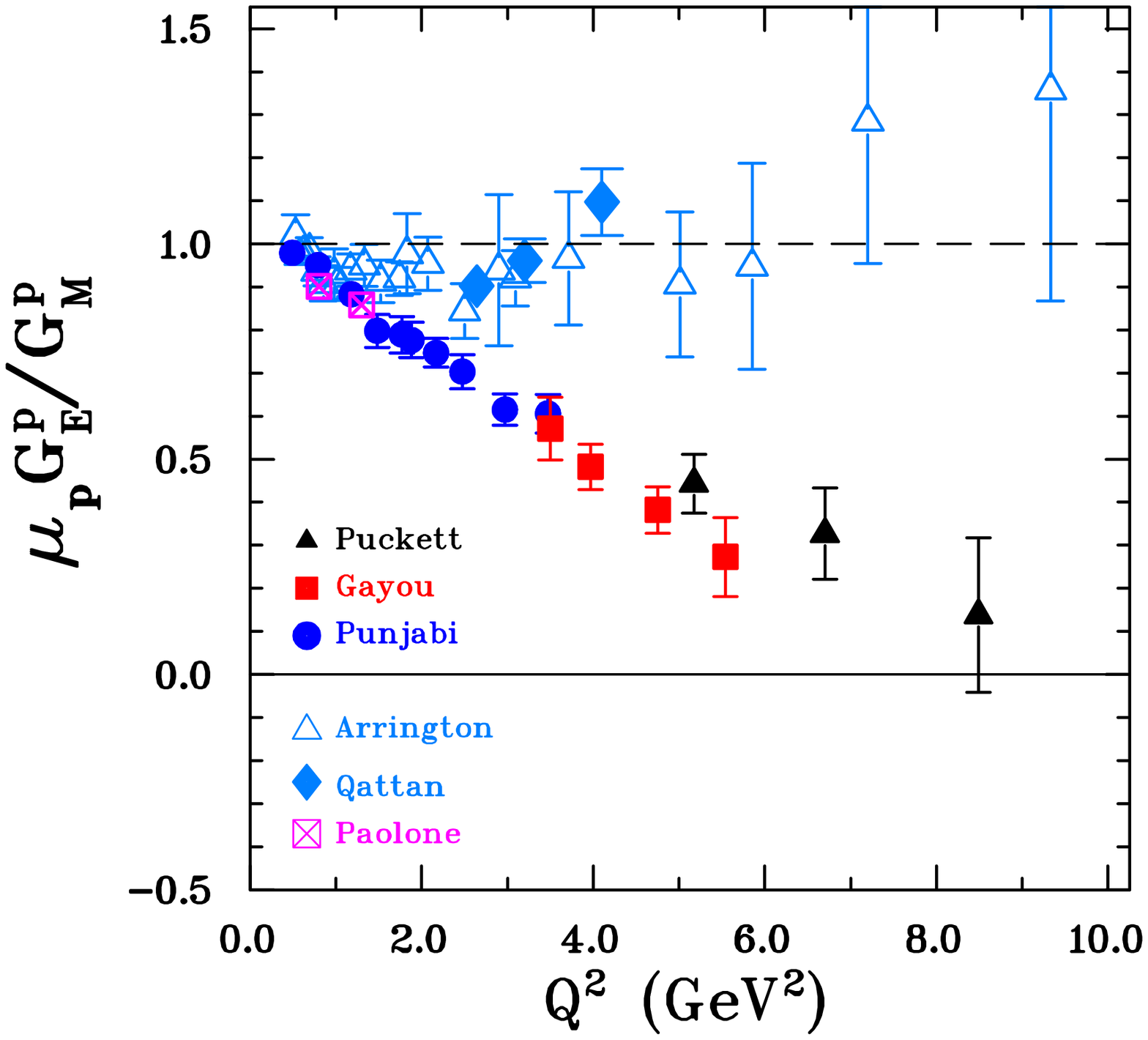,width=0.48\linewidth}
\epsfig{file=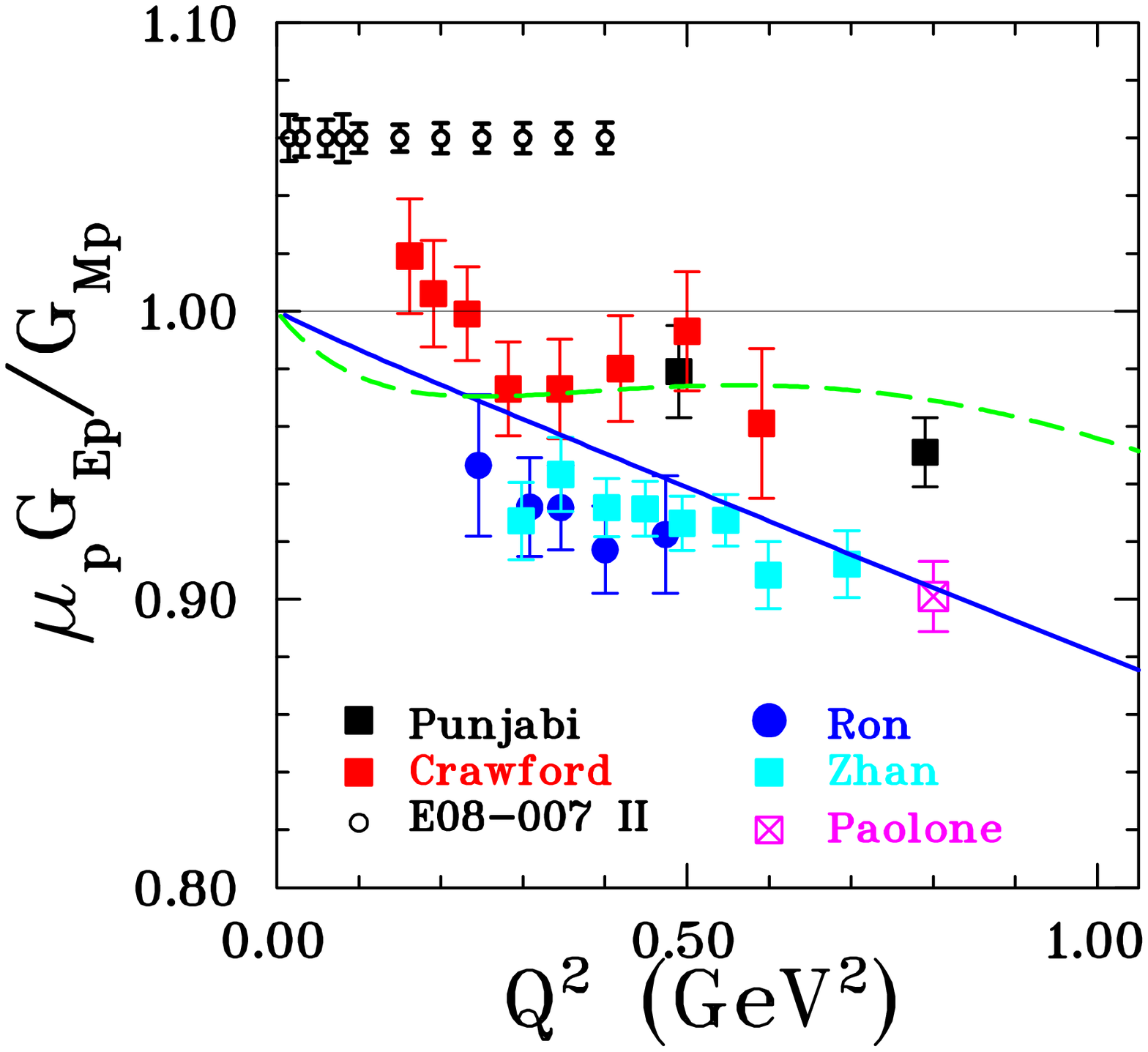,width=0.48\linewidth}
\end{center}
\caption{Left: The ratio $\mugegmp$ from JLab and MIT-Bates (BLAST)
polarization experiments, a global analysis of cross-section
measurements~\cite{arrington04a} (hollow triangles), and the
``Super-Rosenbluth'' results~\cite{qattan05} (solid diamonds). Right: Focus on
the low-$Q^2$ results from polarization measurements~\cite{punjabi05,
crawford07, paolone10, zhan11, ron11}, and projections for the polarized
target measurement of E08-007.  The dashed green line is the fit of
Kelly~\cite{kelly04}, made before most of these data were available, and 
an updated~\cite{zhan11} version of the fit from Ref.~\cite{arrington07c}.}
\label{fig:gepgmp}
\end{figure}

GEp(I) ran in mid-1998 and measured the $\gegmp$ ratio up to
3.5~GeV$^2$~\cite{punjabi05}. The results, shown in Fig.~\ref{fig:gepgmp},
revealed an unexpected decrease of $\gegmp$ with increasing $Q^2$, in
disagreement with scaling.  When GEp(II) extended the measurement of $\gegmp$ to
5.6~GeV$^2$, the ratio was found to continue to decrease linearly with $Q^2$,
to a value of $\mugegmp$=0.28$\pm$0.09 at 5.6~GeV$^2$~\cite{gayou02}. The
GEp(III) measurement extended the measurements up to
8.5~GeV$^2$~\cite{puckett10}, with the data suggesting a slower decrease of
the ratio $\gegmp$ above 5--6~GeV$^2$.

The results from polarization transfer and Rosenbluth extraction of $\gegmp$
are compared in Fig.~\ref{fig:gepgmp}. The discrepancy between the two results
is significant over a wide range of $Q^2$. A detailed reanalysis of the world's
cross-section data~\cite{arrington03a} showed that the apparent discrepancy
between different extractions of $\gep$ (Fig.~\ref{fig:proton_early}) was the
result of neglecting normalization uncertainties when combining data from
different measurements, but that these uncertainties could not account for the
systematic difference between Rosenbluth and polarization-transfer results.

Following the unexpected results of the first polarization-transfer
experiment, a new Rosenbluth separation of unprecedented accuracy was prepared
and the experiment, E01-001, obtained data in 2002. The unique feature of this
``Super-Rosenbluth'' experiment was the detection of the proton, $(e,p)$,
yielding fixed proton momentum for a given $Q^2$, minimal variation of rate
as a function of angle, and smaller
$\varepsilon$-dependent corrections than $(e,e')$. This, combined with smaller
radiative corrections, led to significantly reduced uncertainties in the ratio
$\gegmp$. The results~\cite{qattan05} (solid diamonds in
Fig.~\ref{fig:gepgmp}) cover a $Q^2$-range from 2.6 to 4.1~GeV$^2$, and agree
with the previous Rosenbluth extractions.  At the same time, traditional
Rosenbluth measurements in Hall C~\cite{christy04} provided additional
measurements, albeit with much larger uncertainties (these data are included
in the global analysis). At that point it was clear that there was a
systematic difference between Rosenbluth and recoil-polarization data.

Intensive discussion of various possible explanations followed.  With
the exception of two-photon exchange (TPE) contributions, radiative corrections
were believed to be well understood for the unpolarized case, and small in the
relevant polarization observables~\cite{afanasev01a}. Historically, only
the IR-divergent component of the TPE corrections had been taken into account.
The remaining TPE terms had been evaluated in a soft-photon
approximation~\cite{maximon00}, and found to be small but potentially
important at large $Q^2$-values.  While the TPE contributions are
small, the contribution from $\gep$ decreases rapidly as $Q^2$ increases,
and so the TPE corrections can be important in understanding the discrepancy.  It
was later shown that a $\varepsilon$-dependent correction of $\sim$5\%
could bring the Rosenbluth results in agreement with the polarization
measurements~\cite{guichon03, arrington04a}, while having little impact on
the polarization-transfer results.  The status of these studies and of new
experiments designed to measure two-photon exchange contributions are
discussed in Sec.~\ref{sec:tpe}.

\subsection{Low-$Q^2$ regime}\label{sec:plowq}

The $Q^2$-region below 1~GeV$^2$ is of interest because it covers the range
over which the pion cloud is believed to make a significant contribution to
the electromagnetic structure of the proton and the neutron.  While this
yields a significant
part of the neutron's electric form factor at small $Q^2$, it is also
expected to be important for the form factors of the proton, as discussed for
example by Friedrich and Walcher~\cite{friedrich03}.  Extremely low values of
$Q^2$ are interesting because the form-factor behavior as $Q^2 \rightarrow 0$
is connected to the nucleon charge and magnetic radii.

While the neutron was the focus of early measurements at low $Q^2$, the
demonstration that extremely precise measurements of $\gegmp$ were possible
led to improved low-$Q^2$ measurements of the proton form factor.  In recent
years careful studies of the optical properties of the Hall A HRS
spectrometers have resulted in a significant reduction of the systematic
errors in polarization-transfer data, thus allowing measurements of $\gegmp$
below $\sim$1~GeV$^2$ with a total error of 1\% or less. New data
from Bates~\cite{crawford07}, along with updated results~\cite{ron11} from
JLab `LEDEX' experiment E05-013~\cite{ron07} and preliminary results from the
dedicated E08-007 measurements~\cite{zhan11}, are shown in the right panel of
Fig.~\ref{fig:gepgmp}.  These high precision data do not show any indication 
of structure in the low-$Q^2$ ``pion cloud'' region, although there is at
present a systematic disagreement between the polarized target data from
Bates~\cite{crawford07} and the high precision JLab polarization
data~\cite{ron11,zhan11}.  New results from an extensive set of cross section
measurements at low $Q^2$ from Mainz~\cite{bernauer10} also see a reduction
in $\mugegm$ in this region, consistent with the new JLab data, although
they only provide a fit to the form factors, rather than direct extractions
of $\ge$ and $\gm$.  Phase II of E08-007 will make extremely high precision
measurements using a polarized target.  This will provide another comparison
between the polarized target and recoil polarization measurements, and will
extend the low-$Q^2$ data down to $Q^2=0.015$~GeV$^2$. This will allow for
significantly improved extractions of the magnetic form factor at very low
$Q^2$, where the cross section has greatly reduced sensitivity.

Knowledge of the form factors is also important in the analysis of a range of
other experiments.  Good knowledge of the $e$--$N$ scattering cross section is
crucial to the interpretation of high-precision quasi-elastic scattering
measurements aimed at understanding nucleons in nuclei.  They are also
necessary input to the analysis of parity-violating electron scattering, where
the contributions from strange quarks can be isolated given sufficient
precision on the electromagnetic form factors and the parity-violating
asymmetry.  In addition, a better determination of the charge and magnetic
radii provides input for the hadronic corrections to the hyperfine structure
of hydrogen.

\subsection{Two-photon exchange}\label{sec:tpe}

After the observation of a large discrepancy between Rosenbluth and
polarization measurements of $\mugegm$, two-photon exchange corrections
received a great deal of attention as a possible explanation.  In 2003,
two papers, appearing back-to-back in PRL, shed significant light on
the issue.  The first paper~\cite{guichon03} provided a general formalism for
scattering beyond the Born approximation, while the second~\cite{blunden03}
provided a direct calculation of the TPE correction, including the case of two
hard photons, calculated in a hadronic basis.

The analysis of Guichon and Vanderhaeghen~\cite{guichon03} not only provided
the general formalism, but also demonstrated that relatively small TPE
contributions might bring the
Rosenbluth results into agreement with the new polarization data.  They were
able to resolve the discrepancy with TPE amplitudes at the 2--3\% level,
consistent with the expectation of order $\alpha_{EM}$ corrections for
higher-order electromagnetic diagrams. These TPE corrections, derived based on
a set of very simple assumptions, were, however, at odds with the measured
cross-section ratio of positron-proton and electron-proton
scattering~\cite{arrington04b}. A later analysis~\cite{arrington04d} used
modified assumptions to extract a set of TPE amplitudes of similar magnitude
that could explain the discrepancy between Rosenbluth and
polarization-transfer measurements while being consistent with the comparison
of electron and positron data.  Because the impact of the TPE corrections
is much smaller for polarization transfer measurements, the impact of TPE
on the neutron form factors is also small.  Rosenbluth extractions of
$\gmn$ were not very precise, and all precision extractions of $\gen$ were
based on polarization measurements.  The recent measurement of the
$\varepsilon$ dependence of the polarization observables~\cite{meziane10}
allows a more detailed extraction of the TPE amplitudes~\cite{guttmann10,
borisyuk10}.  There is still model dependence in the $\varepsilon$
dependence of the TPE amplitudes is not fully constrained, but the inclusion
of the polarization observables provides additional important constraints.

Following the initial calculation of TPE corrections in a hadronic model,
several other theoretical calculations of the two hard-photon exchange
contribution were undertaken (see~\cite{carlson07}). A difficulty arises from
the fact that in the intermediate state inside the box diagram, the proton can
be any excited baryon compatible with angular momentum and parity
conservation. One approach, expected to be reliable at lower $Q^2$, has been
to start with a calculation for an unexcited hadron in the intermediate
state~\cite{blunden03,blunden05a}, and then estimate the effect of higher resonance
states~\cite{kondratyuk07}.  Another approach focuses on the high-energy
region~\cite{chen04,afanasev05a}, assuming that the virtual photon interacts with one
valence quark and that the residual system of quarks and gluons is accurately
described by Generalized Parton Distributions (GPDs).  These two approaches
yield qualitatively similar behavior: small corrections to polarization
observables and a small change in the $\varepsilon$-dependence of the cross
section.  This change in the $\varepsilon$-dependence results in a large
correction to the Rosenbluth extraction of $\gep$ at high $Q^2$, where the
initial $\varepsilon$-dependence is extremely small.

While much of the focus has been on improving the calculation of TPE diagrams,
there have also been efforts to improve the treatment of higher-order terms in
the radiative corrections~\cite{maximon00,afanasev02a, bystritskiy07,
afanasev07}.  For example, hard Bremsstrahlung terms yield an additional
$\sim$1\% difference between high and low $\varepsilon$
values~\cite{afanasev07} that is not present
in the soft-photon approximation.  However, as with most of these comparisons,
the result is compared to Mo and Tsai radiative corrections~\cite{mo69}. While
this is the general approach adopted in nearly all experimental extractions of
the cross sections, many analyses have applied improved corrections, e.g. for
multi-photon bremsstrahlung, and further work is required to determine to what
extent the corrections actually applied to the data differ from the new
calculations.

While the reduced cross section must depend linearly on $\varepsilon$ in
the Born approximation, TPE contributions will introduce a curvature in the
$\varepsilon$-dependence of the cross section (in the Rosenbluth procedure).  A
global analysis of Rosenbluth measurements set tight limits on the non-linear
contributions to the $\varepsilon$-dependence of the cross
section~\cite{tvaskis06}, but these limits are not tight enough to rule out
the non-linear contributions predicted by most TPE calculations.  New
measurements of the kinematical dependence of the $\gegmp$ ratio, obtained in
recoil polarization at a constant $Q^2$ of 2.48~GeV$^2$ in Hall C (E04-019),
show no $\varepsilon$-dependence within the statistical
uncertainty~\cite{pentchev08}, supporting the calculations that conclude that
polarization measurements are not significantly affected by TPE. The most
direct test of TPE contributions, the comparison of positron and electron
scattering, yields evidence for a non-zero TPE contribution at the 3$\sigma$
level~\cite{arrington04b}.  At the present time there is no definitive
experimental evidence that TPE contributions are the major cause of the
difference between Rosenbluth and recoil-polarization extractions of the
proton form factor, although preliminary results from positron-electron
comparisons at Novosibirsk~\cite{vepp_proposal} yield an excess positron
cross section consistent with TPE calculations, and data taking has begun
for E07-005, a much more extensive set of comparisons in Hall B.

There is an active program at JLab to determine if TPE corrections do in fact
fully explain the discrepancy, and to map out their impact on both the cross
section and polarization observables.  In Hall C, E05-017 extended the
high-precision ``Super-Rosenbluth'' measurements, covering a large range in
$Q^2$ and $\varepsilon$ to better map out the Rosenbluth-polarization
difference and to improve the limits on non-linear contributions by a factor
of two or more over a wide range in $Q^2$.  The discrepancy has motivated
experiments to compare positron and electron scattering at Novosibirsk,
DESY, and Jefferson Lab~\cite{arrington09b}. Data taking has begun for E07-005,
which will provide a new comparison of the $e^+$--$p$ and $e^-$--$p$ cross
sections over a broad kinematic range.  This experiment is running in Hall B
using a tertiary beam of both positrons and electrons over a broad range of
energies.  This will allow a quantitative study of TPE contributions over a
range of $\varepsilon$ for $Q^2$ up to 2--3~GeV$^2$, and will directly
determine whether TPE corrections fully explain the discrepancy between
Rosenbluth and polarization measurements in this region.  While $\gep$ is a
unique case where these small corrections have a large impact on the
interpretation of the results, TPE effects contribute to all
electron-scattering measurements. A stringent test of the calculations is
important to ensure that these corrections are well understood for future
measurements that aim for extremely high precision in a variety of reactions.

\section{Neutron Form Factors}\label{sec:neutron}

For the neutron, form-factor extractions using the Rosenbluth technique
were limited by the small size of the electric form factor and the need to
measure $e$--$n$ scattering using quasi-elastic scattering from deuteron targets.
As discussed in Sec.~\ref{sec:history}, it was known that $\gen$ was positive,
but there was a factor of two uncertainty in the values derived from elastic
electron--deuteron scattering.  While the situation for $\gmn$ was better,
both the precision and $Q^2$-range were limited compared to measurements on
the proton.

The availability of high-polarization beams, effective polarized-neutron
targets, and large neutron detectors and recoil polarimeters made it possible
to dramatically improve our knowledge of the neutron form factors.  Experiments
at other labs had demonstrated the feasibility of such measurements, and begun
the process of validating the techniques through comparisons of measurements
utilizing different techniques or different target nuclei.  The JLab program
added to these studies by expanding precision measurements of $\gmn$ above
$Q^2$=1~GeV$^2$ and by providing essentially all of the direct measurements of
$\gen$ above 0.8~GeV$^2$.

\subsection{Neutron magnetic form factor}

\begin{figure}[h]
\begin{center}
\epsfig{file=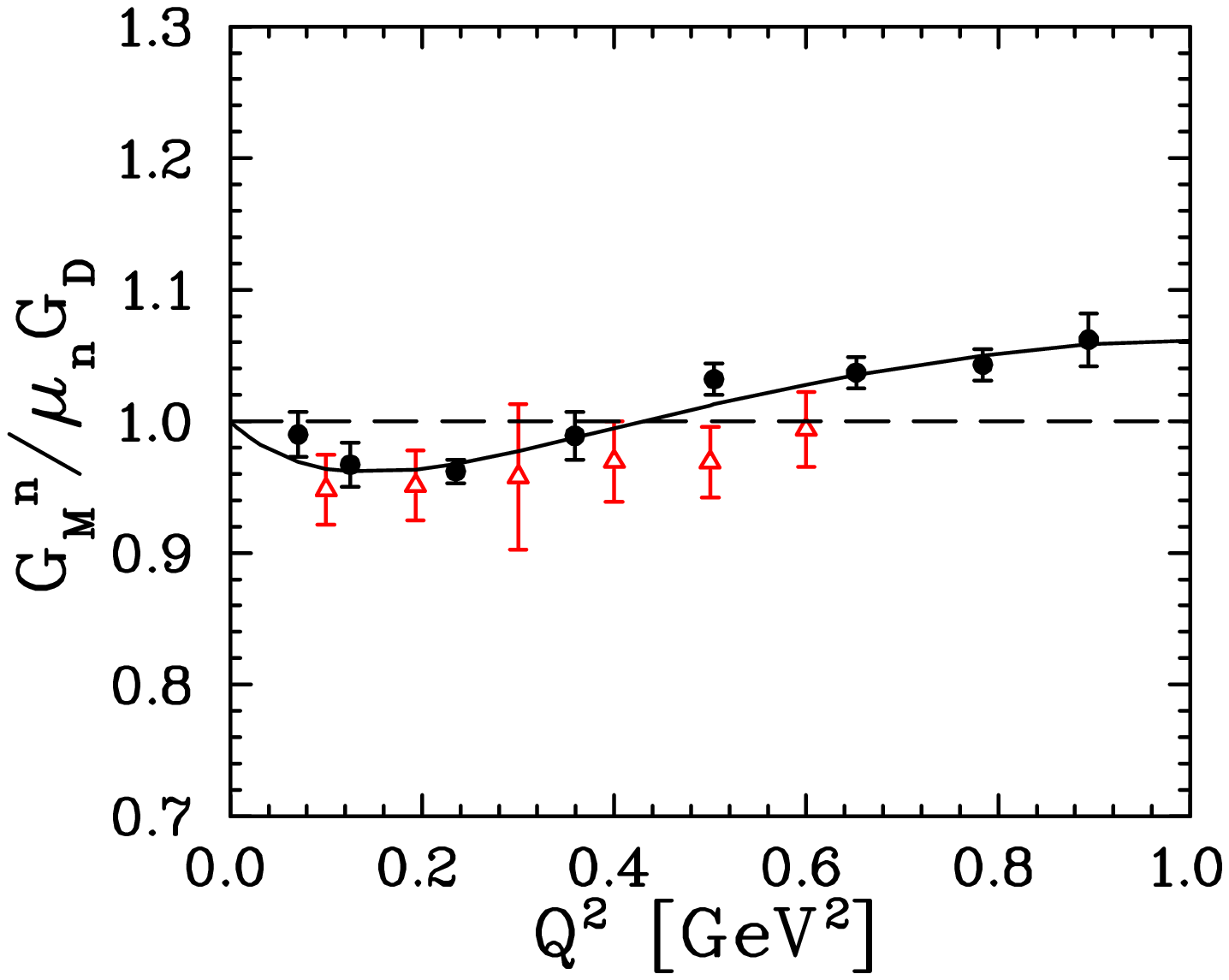,width=0.48\linewidth,height=0.355\linewidth}
\epsfig{file=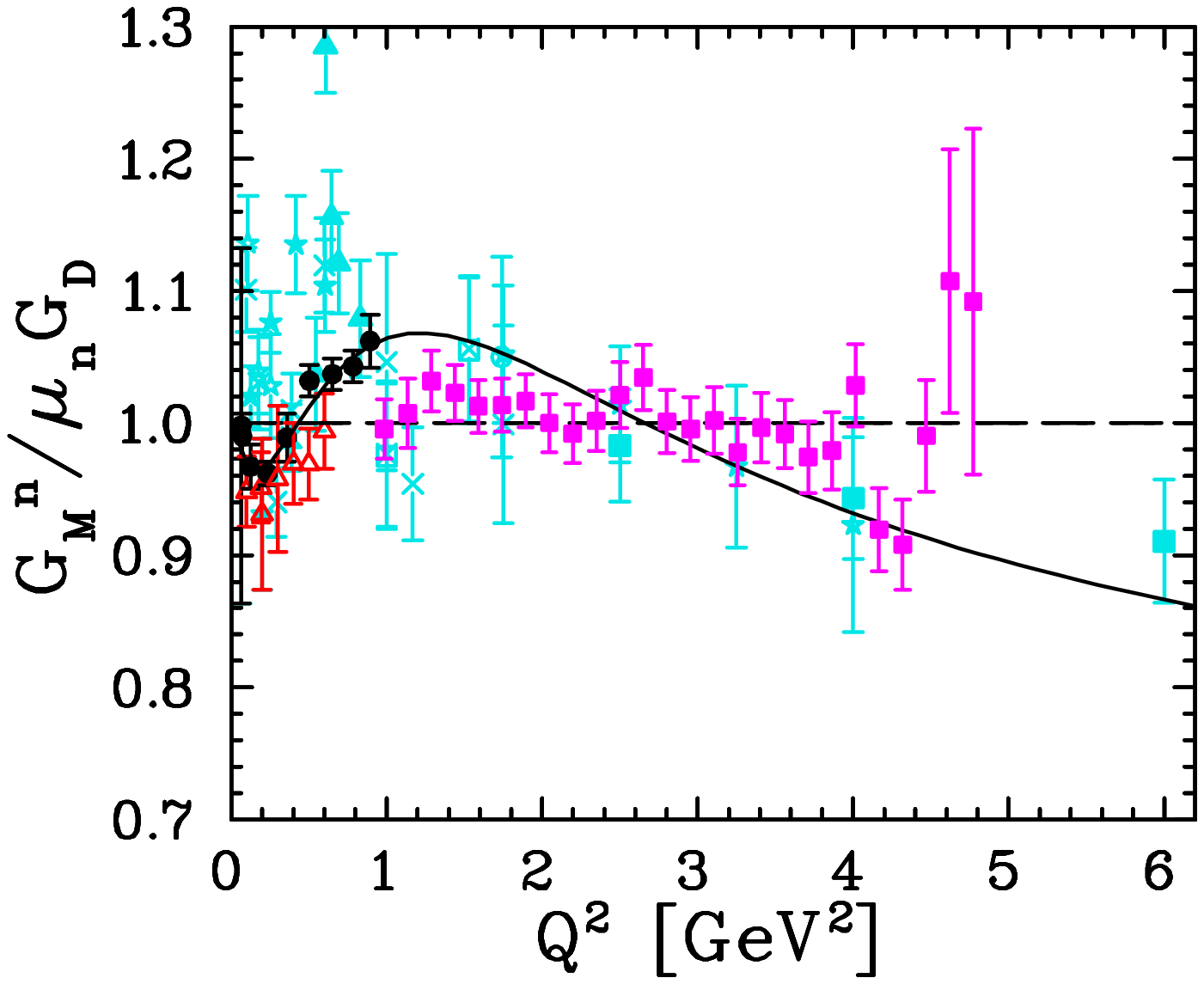,width=0.48\linewidth,height=0.355\linewidth}
\end{center}
\caption[]{Current status of extractions of $\gmn/\mu_n/G_D$.  The left figure
shows only the high precision, low $Q^2$ JLab polarization~\cite{anderson07}
(open triangle) and Mainz ratio~\cite{anklin98,kubon02} (closed circles)
measurements, while the right figure shows the full data set, including
the CLAS ratio~\cite{lachniet08} measurements (solid squares).}
\label{fig:neutron_gmn}
\end{figure}

Figure~\ref{fig:neutron_gmn} shows the present status of measurements of the
neutron magnetic form factor.  In the high-$Q^2$ regime,
E94-017~\cite{lachniet08} completed a study of $\gmn$ at \q2 up to
$\sim$5~\gev2 in 2000 by measuring the exclusive neutron/proton cross-section
ratio from the deuteron with the CLAS detector in Hall B. The ratio of the
$^2$H$(e,e'n)$ and $^2$H$(e,e'p)$ reactions in quasi-elastic kinematics is
approximately equal to the ratio of elastic scattering off a free neutron and
proton, respectively:
\begin{equation} 
R_D = \frac{\frac{d\sigma}{d\Omega}[^2\mbox{H}(e,e'n)_{QE}]}
           {\frac{d\sigma}{d\Omega}[^2\mbox{H}(e,e'p)_{QE}]}
= a \cdot R_{free} 
= a \cdot \frac{\tau(\gmn)^2+\varepsilon(\gen)^2}{\tau(\gmp)^2+\varepsilon(\gep)^2}
\end{equation}
The correction factor $a$ is close to unity for quasi-elastic kinematics and
larger \q2-values and can be accurately calculated~\cite{jeschonnek00} as a function of \q2 and
$\theta_{pq}$, the angle between the momentum transfer and the knocked-out
nucleon in the center of mass, using standard models for the deuteron. The
value of $\gmn$ is then determined from the measured value of $R_D$ and the
known values of $\gmp$ and $\gep$, with very small corrections due to the lack
of knowledge of $\gen$.  The cross sections for the $^2$H$(e,e'n)$ and $^2$H$(e,e'p)$ 
reactions were
measured simultaneously. Measuring the ratio $R_D$ reduces or completely
eliminates several systematic uncertainties, such as those from final-state
interactions in the deuteron, from knowledge of the luminosity and from
radiative corrections. The remaining major uncertainty was due to the neutron
detector efficiency, which was measured through pion electroproduction off a
hydrogen target. These efficiency measurements were performed simultaneously
with the primary deuteron measurement, by positioning a hydrogen target
upstream of the deuterium target. The experiment took data at two different
beam energies, and for each energy independently analyzed neutrons detected in
time-of-flight scintillators and neutrons detected in the calorimeter, yielding
four essentially independent but overlapping measurements~\cite{lachniet08}.

Experiment E95-001~\cite{xu00, xu03, anderson07} determined $\gmn$ in 1999 at
\q2-values smaller than 1~\gev2 by measuring the beam asymmetry in inclusive
quasi-elastic scattering of electrons off a polarized \3he target in Hall A.
The polarized $^3$He target system in Hall A utilizes the spin exchange
between optically pumped alkali-metal vapor and noble-gas nuclei to produce an
ensemble of spin-polarized $^3$He nuclei. Polarized $^3$He nuclei serve as an
effective polarized neutron target, because their ground state is dominated by
a spatially-symmetric $s$-state in which the spins of the protons cancel. The
central feature of the target system is a sealed glass cell, that contains
an admixture of $^3$He gas at a pressure of 0.7 MPa and an alkali-metal vapor.
These cells have two chambers, a heated upper chamber in which the spin
exchange takes place and a lower one, through which the electron beam passes.
Lower cells with lengths from 20 to 40 cm have been used, corresponding to a
target thickness of up to $1 \cdot 10^{22}$ nuclei/cm$^2$.  In earlier versions
of the target only rubidium was used as the spin-exchange medium, resulting in
an in-beam polarization of up to $\sim$40\% at a beam current of up to $\sim$10~$\mu$A.  
A detailed description of the
polarized $^3$He target system can be found in Ref.~\cite{alcorn04}.

The beam-target asymmetry is:
\begin{equation} 
A = - \frac{(\cos \theta^* v_{T'}R_{T'}+2 \sin \theta^* \cos \phi^* v_{TL'}R_{TL'})}
{ v_L R_L + v_T R_T},
\end{equation}
where $\theta^*$ and $\phi^*$ are the polar and azimuthal target polarization
angles with respect to $\vec{q}$, $R_i$ denote various nucleon response
functions, and $v_i$ the corresponding kinematic factors. By orienting the
target polarization parallel to $\vec{q}$, one measures the ratio of
$R_{T^\prime}$ to the unpolarized cross section. In quasi-elastic kinematics
$R_{T^\prime}$ is dominantly sensitive to $(\gmn)^2$:
\begin{equation} 
R_{T^\prime} \propto P_n (G_M^n)^2 + P_p (G_M^p)^2,
\end{equation}
where $P_n$ and $P_p$ denote the effective polarizations of the neutron and
the proton, respectively. The extraction of $\gmn$ requires an iterative
process since the asymmetry depends on both $R_{T^\prime}$ and the
unpolarized cross section, which also depends strongly on $\gmn$.
In addition, corrections for the nuclear medium~\cite{golak01} are necessary
to take into account effects of final-state interactions and meson-exchange
currents. Such corrections are calculable non-relativistically at low
\q2~\cite{xu00}. At intermediate $Q^2$, relativistic effects have to be taken
into account, making calculations much more difficult. However, there the size
of the corrections is expected to be small and they have been neglected in the
analysis~\cite{xu03}. Both measurements are in good agreement in the overlap
region (Fig.~\ref{fig:neutron_gmn}) and with the CLAS data in the larger
\q2-range, but disagree significantly from earlier Mainz
measurements~\cite{anklin98, kubon02}, that used the same ratio method as the
CLAS measurement, but with an off-site neutron detector calibration.

\subsection{Neutron charge form factor}

The same techniques that have been used to measure $\gegmp$ can also be 
applied to measure $\gegmn$, except that one measures scattering from
neutrons in $^2$H or $^3$He. In the past decade, a series of JLab beam-asymmetry 
measurements of neutron knock-out from a polarized target  or
studies of polarization transfer have provided accurate data on $\gen$. The
first such measurements at JLab were carried out in Hall C.

Arnold, Carlson and Gross~\cite{arnold81} were the first to show that the
measurement of the up-down asymmetry in a neutron polarimeter after the
spin of the knocked-out neutron has been precessed by a vertical dipole field yields access to
the ratio $\gegmn$.  This was the technique used by experiment E93-038, which
in 2000 used a large scintillator neutron polarimeter to determine $\gen$ at
\q2-values of 0.45, 1.13 and 1.45~\gev2~\cite{madey03}. The neutron
polarimeter consisted of a large dipole magnet with a vertically oriented
field, an active analyzer (preceded and followed by a veto/tagger), and a top
and bottom array of scintillators. The dipole magnet precesses the spin of the
neutrons through an angle $\chi$ in the horizontal plane and sweeps protons
and other charged particles out of the acceptance. The active analyzer
consists of twenty 100$\times$10$\times$10~cm$^3$ scintillators. The long axes
were oriented horizontally, perpendicular to the central flight path, stacked
vertically into four layers of five detectors. The up/down rear arrays of
scintillators each consisted of 12 detectors stacked in three horizontal
layers with the long axes of the scintillators oriented parallel to the flight
path of the neutrons. The up-down scattering asymmetry measured in this rear
array is proportional to the projection of a recoil polarization on to a
horizontally-oriented sideways axis. Additional elements of this polarimeter
were a lead curtain, veto taggers and extensive steel and concrete shielding
surrounding the scintillator detectors~\cite{plaster06}.

The precession through an angle $\chi$ results in a scattering asymmetry
$\xi(\chi)$
\begin{equation}
\xi(\chi) = A_y P^n \sin (\chi - \delta)
\end{equation}
The ratio $\gegmn$ can be determined from the value
of the precession angle where the polarization asymmetry is observed
to be zero:
\begin{equation} 
 \frac{\gen}{\gmn} = - \sqrt{\frac{\tau(1 + \varepsilon)}{2 \varepsilon}} \tan \delta .
\end{equation}
By determining a so-called cross-ratio of the up-down asymmetry in the
polarimeter for both values of the beam helicity the result becomes
independent of both the target luminosity and the polarimeter efficiency.
Corrections for charge-exchange reactions in the lead curtain were determined
to be $\approx$3\% based on measurements with a liquid hydrogen target and
detailed Monte-Carlo simulations.

E93-026 ran in 2001 and used a deuterated ammonia (ND$_3$) target as an
effective polarized neutron target to measure
$\gen$ at \q2 of 0.5 and 1.0~\gev2~\cite{warren04}. The solid polarized
target used the dynamical nuclear polarization
technique~\cite{crabb95, averett99} to reach an in-beam polarization of
$\sim$24\% at electron beam intensities of up to 100 nA. Ammonia ND$_3$
granules, doped by radiation damage with a small concentration of free
radicals, were immersed in liquid helium. Because the occupation of the
magnetic substates in the radicals follow a Boltzmann distribution, the free
electrons are polarized to more than 99\% in a $\sim$5~T field generated by a
pair of superconducting coils and at a $\sim$1~K temperature. A radiofrequency
field is then applied to induce transitions to states with a preferred nuclear
spin orientation. Because the relaxation time of the electrons is much shorter
than that of the nuclei, polarized nuclei are accumulated. The electron beam
had to be rastered uniformly in a 1 cm radius to minimize local heating and
depolarization. A two-magnet chicane compensated for the deflection of the
electron beam by the target field. The scattered electrons were detected in
the HMS detector and the knocked-out neutrons in a 160$\times$160 cm$^2$ large
scintillator detector, six planes thick, preceded by two planes of thin
scintillators serving as veto detectors of charged particles, shielded by a
2.5 cm thick lead curtain from direct gamma rays originating in the target.
The measured beam-target asymmetry $A_{en}^V$, with the target polarization
vector in the scattering plane and perpendicular to the momentum transfer
vector, can be directly related to the ratio $\gegmn$
\begin{equation} 
A_{en}^V = - \frac{2 \sqrt {\tau (\tau + 1)} \tan (\theta_e/2) \gegmn}
{(\gegmn)^2 + \tau / \epsilon}
\end{equation}
Corrections for charge-exchange reactions in the target material and in the
lead curtain were determined by a Monte-Carlo simulation.

\begin{figure}[h]
\begin{center}
\epsfig{file=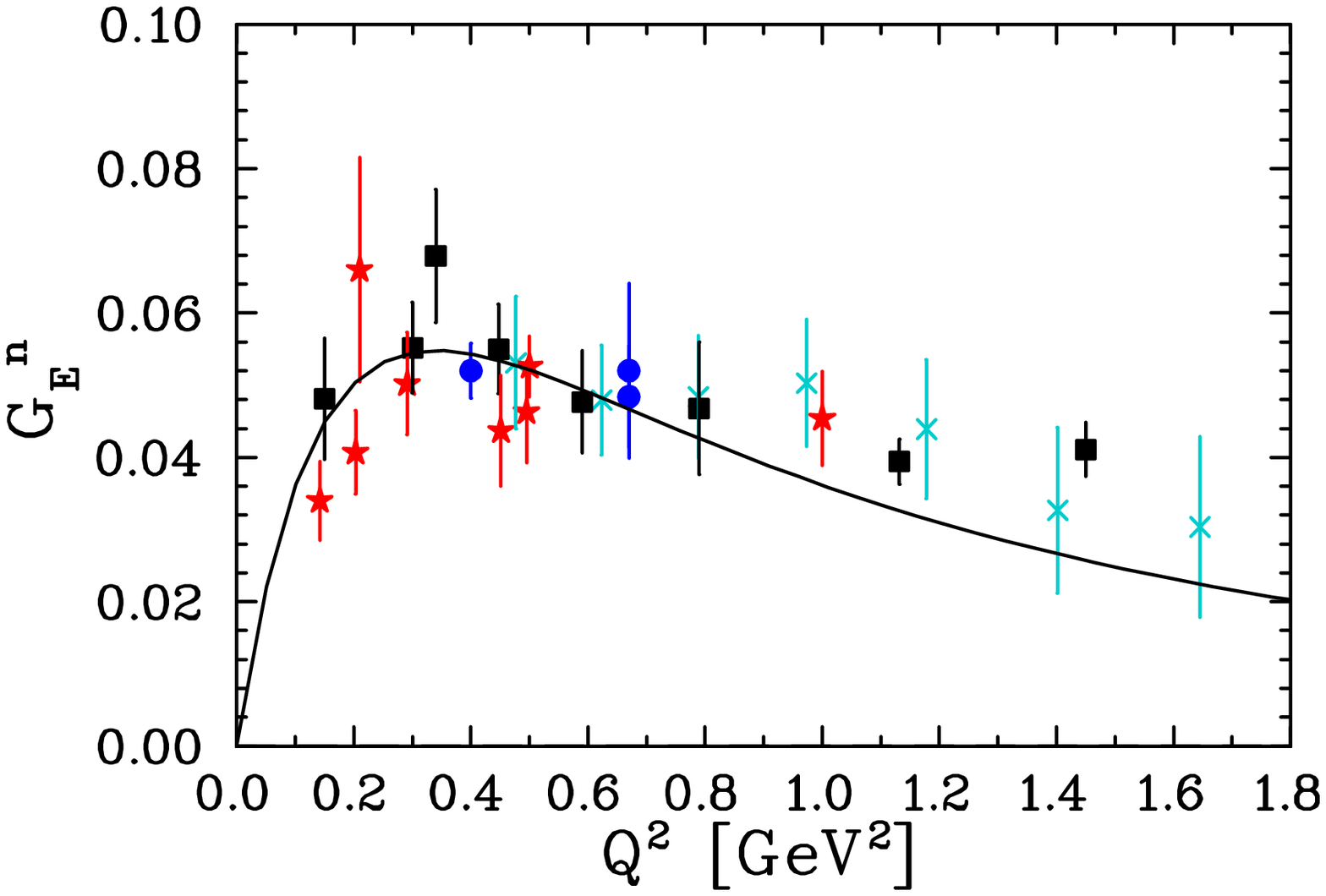,width=0.48\linewidth}
\epsfig{file=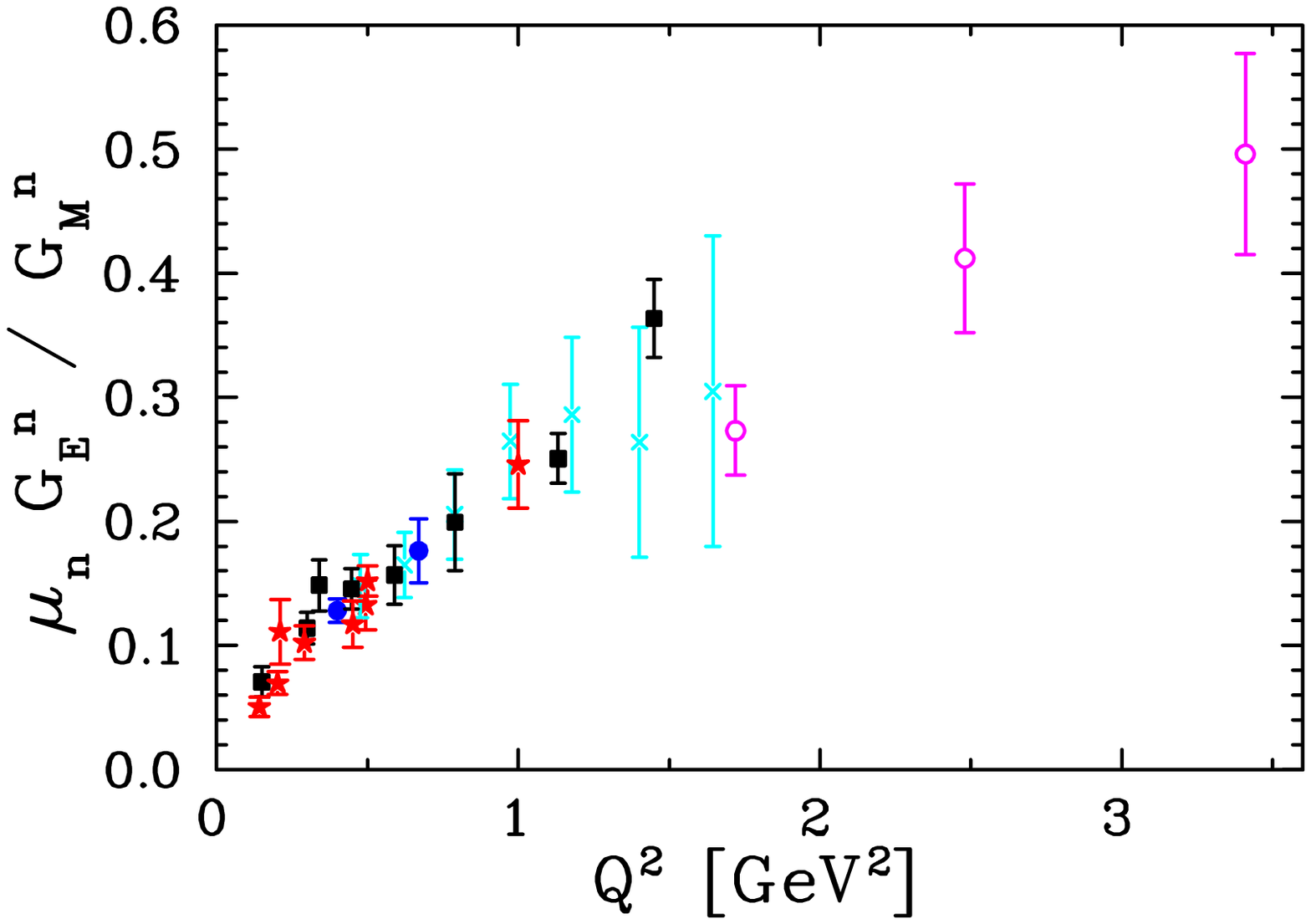,width=0.48\linewidth}
\end{center}
\caption[]{Current status of extractions of $\gen$.  Squares are extractions
from recoil-polarization measurements on $^2$H, stars (circles) are from
cross-section asymmetries from  polarized $^2$H ($^3$He) targets.  The crosses
are from a modern analysis of electron-deuteron elastic
scattering~\cite{schiavilla01}, and the curve represents the Galster fit to
$\gen$. Data sets are the same as shown in Ref.~\cite{arrington07a} with the
addition of new results from BLAST~\cite{geis08} (red stars) and the high-$Q^2$
results from E02-013 (hollow circles).}
\label{fig:neutron_gen}
\end{figure}

Figure~\ref{fig:neutron_gen} shows the combined results from all of the
high-precision extractions of $\gen$.  Results from recoil-polarization
measurements on $^2$H and cross-section asymmetry measurements from polarized
$^2$H and $^3$He targets are in good agreement.  The limited figure of merit
of the polarized $^2$H targets, due to restrictions in the polarization, the
dilution factor, and the beam current, inhibited its use at higher
$Q^2$-values.

Experiment E02-013 ran in Hall A in 2006 with a polarized \3he target, a
large-acceptance neutron detector and a large-acceptance electron spectrometer
(``BigBite''). The target utilized hybrid optical pumping in which a mixture of
rubidium and potassium is used to enhance the spin exchange, resulting in an
in-beam polarization of $\sim$50\%. The BigBite spectrometer consisted of a
large dipole magnet with an angular acceptance of close to 100 msr and a
momentum acceptance of 90\%. Its detector package contained two planes of
drift chambers and a lead-glass shower calorimeter. A special target holding
field magnet was designed that also provided the required magnetic shielding
of the target cell from the fringe field of BigBite. The neutron detector was
the largest dedicated neutron detector ever built. It had an active area of
1.6$\times$5~m$^2$, consisting of 250 scintillators stacked in 7 planes that
were interspersed with a 2.5 cm thick iron conversion plane and preceded by
7.5 cm of lead and iron shielding and 2 veto planes made from 200 thin
scintillator bars. The use of these three novel devices at a luminosity of $3
\cdot 10^{37}$ cm$^{-2}$s$^{-1}$ made it possible to extend the $\gen$ data set
to 3.4 \gev2, as shown in the right panel of Fig.~\ref{fig:neutron_gen}. Charge-exchange corrections were determined by taking data, in
addition to $^3$He, on a number of targets with different N/Z ratios, such as
hydrogen, deuterium and $^{12}$C. Nuclear-medium effects, including pion
production, were determined through Glauber-type calculations.

\section{Impact of the Jefferson Lab Program}

The dramatically improved data set that has become available in the last decade,
has had a transformative effect on the study of nucleon form factors.  While
the issue of two-photon exchange corrections led to a brief period of
uncertainty, it soon became clear that the surprising new results on $\gegmp$
were correct and that our textbook picture of the proton form factors would
have to be revised.  The impact of these new experimental results was
magnified by the parallel developments on the theory side, in particular
several attempts to learn more about the internal sub-structure of the nucleon
within the framework of Generalized Parton Distributions (GPDs).

The most dramatic new result was the fall-off of the ratio $\gegm$ for the\
proton.  One question raised by this striking behavior was whether $\gep$ may
have a zero crossing at high $Q^2$.  This possibility was considered rather
exotic when the high-$Q^2$ polarization results were first available, as the
idea that $\gep$ and $\gmp$ both followed the dipole form and decreased
monotonically towards zero had become common wisdom based on the earlier
Rosenbluth measurements. However, there is nothing  unusual
about a zero crossing in $\gep$.  In fact, Dombey stated in a 1969 review
article~\cite{dombey69} that ``As $\ge = F_1 - \tau F_2$, it is \textit{a priori} quite
likely that $\ge$ becomes negative for large values of [$Q^2$]''. 
The recent GEp(III) experiment~\cite{puckett10} shows a decrease in the fall
of $\gegmp$ with $Q^2$, suggesting that a zero crossing, if it occurs, is at
higher $Q^2$ than suggested by the earlier measurements. Results from the new
$\gen$ experiment~\cite{riordan10} suggest that $\gen$ falls less rapidly,
roughly following the dipole form for $Q^2$ from 1.5 to 3.4~GeV$^2$.

\begin{figure}[hbt]
\begin{center}
\epsfig{file=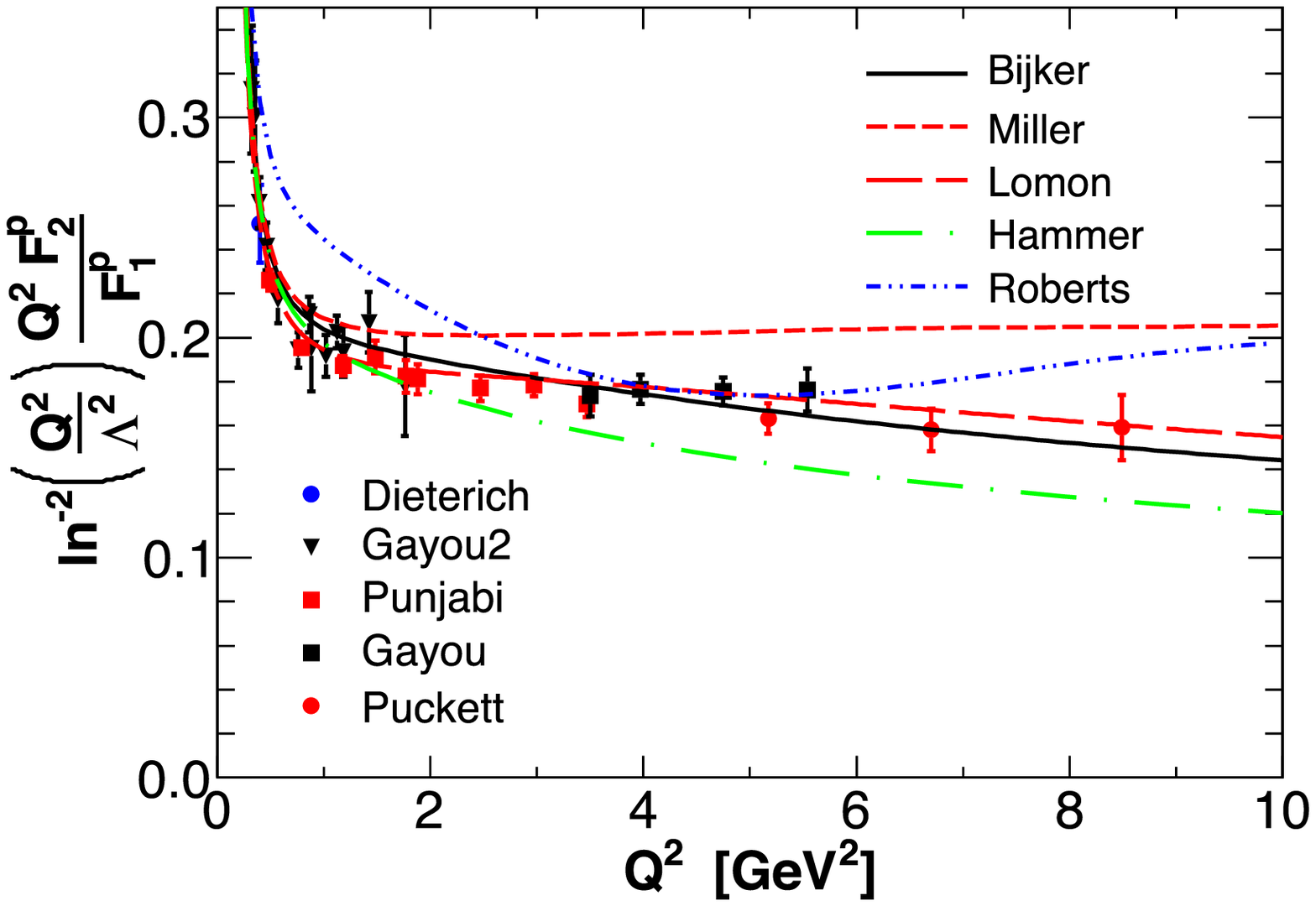,width=0.49\linewidth,height=2.4in}
\epsfig{file=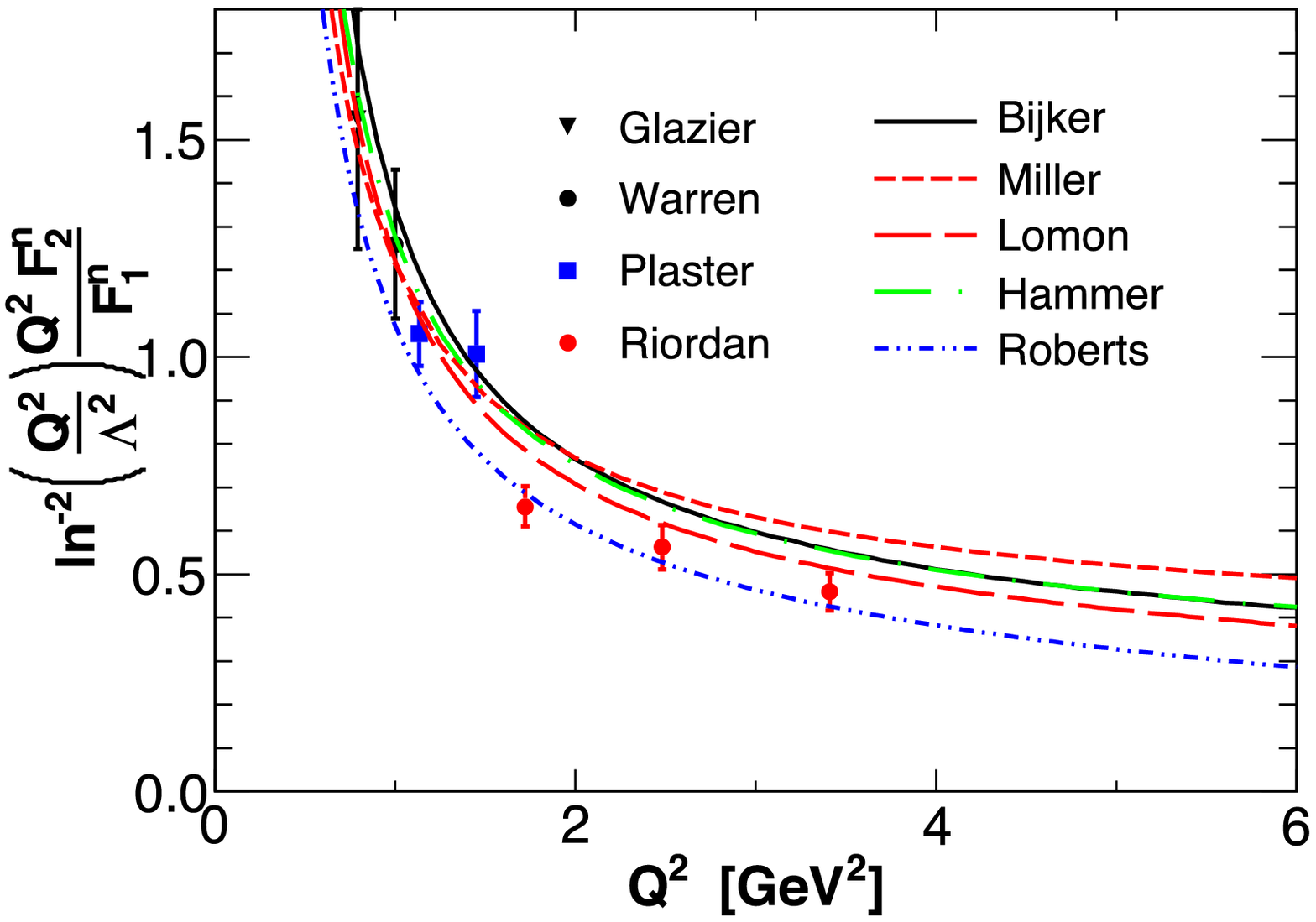,width=0.49\linewidth,height=2.4in}
\end{center}
\caption{The $F_2/F_1$ ratio for the proton and neutron, after applying
a logarithmic scaling correction from~\cite{belitsky03}, with $\Lambda=300$~MeV.
The proton (left) is consistent with this logarithmic scaling above 1--2~GeV$^2$,
while the preliminary neutron results, extending up to $Q^2=3.4$~GeV$^2$, do
not yet show the same scaling behavior.}
\label{fig:scaling}
\end{figure}

The expectation for high $Q^2$ was that the form factors would behave according
to the leading term in perturbative QCD (pQCD).  This led to the expectation
that the ratio $\gegm$ would become independent of $Q^2$ at large $Q^2$-values
(corresponding to $F_1/F_2 \propto Q^2$), as suggested by the older Rosenbluth
results.  The pQCD predictions were reexamined, and while the leading power
behavior yields a constant ratio, there are terms that give a logarithmic $Q^2$-dependence
 that can be large even at the highest $Q^2$-values of
existing data~\cite{belitsky03}.  Figure~\ref{fig:scaling} shows scaling for
the proton and neutron, along with a selection of calculations.  

Miller's calculation~\cite{miller02b} is based on an extension of the cloudy
bag model, in which three relativistically moving constituent quarks are
surrounded by a pion cloud.  Roberts~\cite{cloet08} solves a Poincare
covariant Faddeev equation for dressed quarks in which correlations between
those are expressed via diquarks. The other three calculations use different
extensions of the Vector Meson Dominance model where the scattering amplitude
is expressed in a bare-nucleon form factor, multiplied by the amplitude of the
photon interaction with a vector meson.  In Bijker and Iachello's
model~\cite{bijker04}, the virtual photon is assumed to couple to the assumed
intrinsic structure of the quarks and one of three vector mesons ($\rho$,
$\omega$ and $\phi$).  By adding more parameters, such as the width of the
$\rho$-meson and the masses of heavier vector mesons, Lomon~\cite{lomon02,lomon06}
succeeded in describing all EMFF data. Hammer and Meissner~\cite{hammer04b}
included the isovector $\pi \pi$ channel through dispersion relations.  Because
the VMD models require a significant number of parameters to provide good fits
to the data, they are not expected to have significant predictive power.

While the proton data are consistent with the modified scaling of
Ref.~\cite{belitsky03}, the neutron data do not show this scaling, and it has
been suggested~\cite{arrington07a} that the non-perturbative mass scale
required for the proton indicates that the perturbative prediction is not
applicable in this $Q^2$-range.  These logarithmic terms are connected to
spin-flip contributions, which for nearly massless quarks must come from the
orbital angular momentum of the quarks.  Quark orbital angular momentum is an
important feature in many of the nucleon models that show $\gegmp$ decreasing
with $Q^2$.

\begin{figure}[hbt]
\begin{center}
\epsfig{file=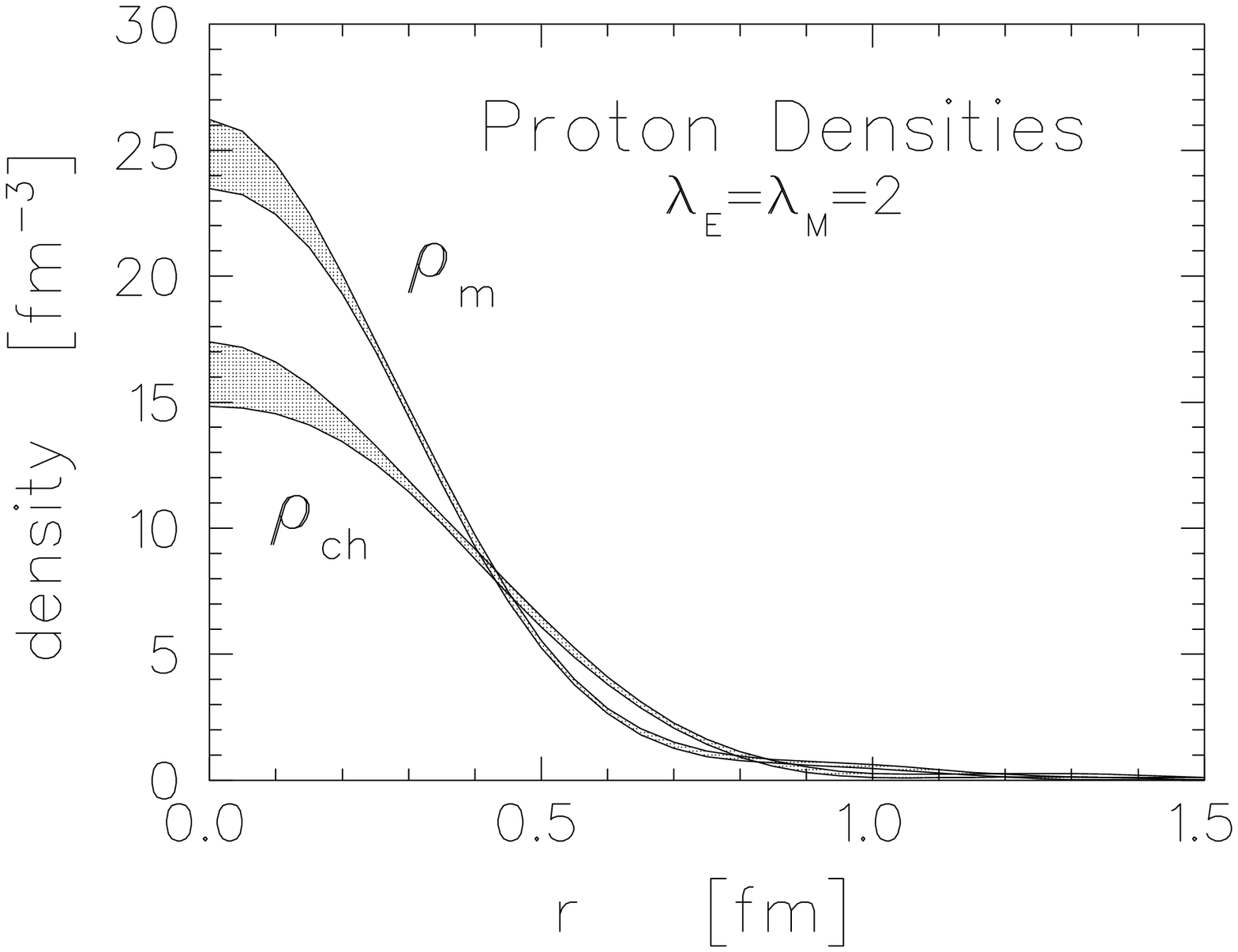,height=0.53\linewidth,width=0.45\linewidth,angle=90}
\epsfig{file=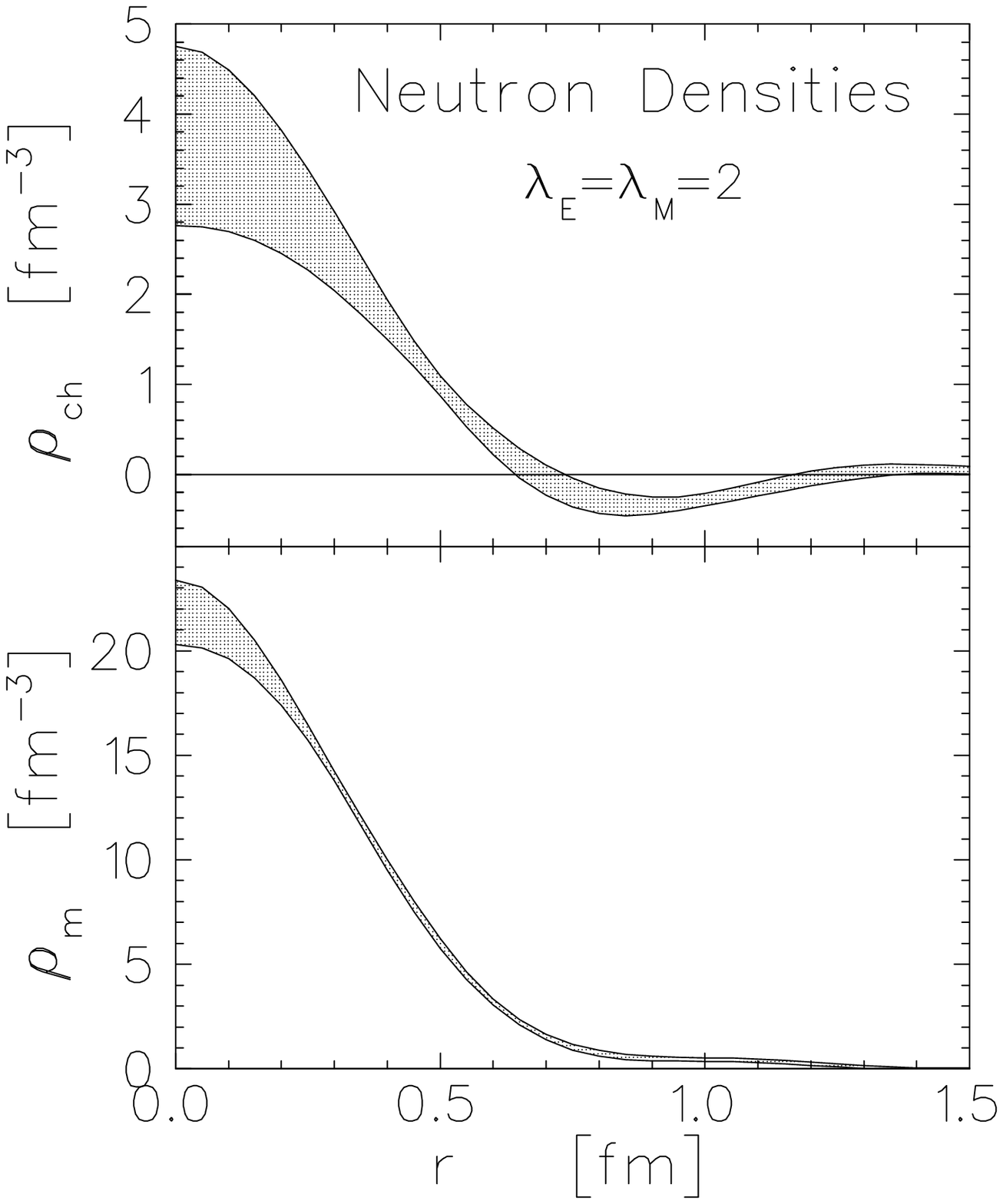,width=0.44\linewidth,height=0.42\linewidth}

\end{center}
\caption{Charge and magnetization densities for the proton and
neutron.  Figures taken from Ref.~\cite{kelly02}.}
\label{fig:densities}
\end{figure}

In the Born approximation, the form factors are Fourier transforms of the
charge and magnetization distributions in the Breit (center-of-momentum)
frame. The extraction of those distributions in the rest frame requires
relativistic boost corrections, which scale as $Q^2/m^2$.  At low $Q^2$, the
mass of the constituent quarks has evolved to $\sim$400 MeV, so the
corrections are expected to be relatively small for very low $Q^2$-values.
Kelly~\cite{kelly02} used a simple model for these boost corrections in
obtaining the results shown in Fig.~\ref{fig:densities} for the charge and
magnetization distributions of the proton and the neutron.  For the proton,
the central magnetization density is 50\% larger than the central charge
density, as a result of its sharper drop-off.  For the neutron, there is a
positive central charge distribution with an extended negative tail, strongly
supporting the picture that the neutron has a ($p,\pi ^-$) component in its
wave function.

\begin{figure}[hbt]
\begin{center}
\epsfig{file=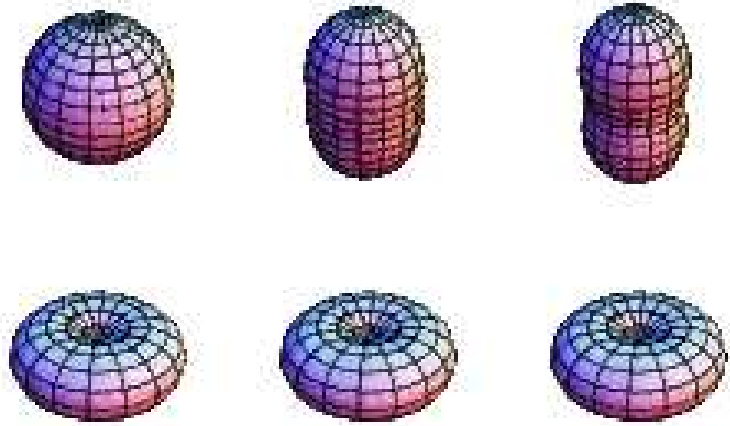,width=0.60\linewidth} \\
\end{center}
\caption{Visualizations of the spatial quark distributions in the
proton~\cite{miller03}.  The top (bottom) distributions are for quark spin
parallel (anti-parallel) to the proton spin.  Contours of constant density are
shown for quark momenta 0, 1, and 2 GeV/c (from left to right).}
\label{fig:miller}
\end{figure}

In parallel with the improvements in the experimental techniques, new tools
were also being developed to allow for the extraction of additional information
about the structure of the nucleon.  The development of the framework of
Generalized Parton Distributions (GPDs) led to new approaches that go beyond
the traditional one-dimensional pictures of the spatial charge distributions
and look into correlations in the momentum, space, and spin structure of the
quarks in the nucleon.  Several groups looked at isolating the spatial
distribution for low and high momentum quarks, or for the quarks with spins
parallel, anti-parallel, or transverse to the nucleon spin~\cite{miller03,
belitsky04}, as illustrated in Fig.~\ref{fig:miller}.  More recently,
both the charge and magnetization densities, along with their uncertainties,
has been extracted~\cite{venkat11}. The impact of these data on constraining
GPDs, and broader discussion of the interpretation of nucleons in terms of the
GPDs is included in a later article in this collection~\cite{hyde11}.

The approaches built on GPDs provided more detailed information on nucleon
sub-structure, but significant modeling is required to build GPDs from the
constraints provided mainly by form factors and structure functions.  Related
studies were performed that led to the development of a model-independent
procedure to extract information on the spatial distributions that depend only
on the form factors as input. Miller~\cite{miller07} showed that in the
infinite momentum frame (IMF), the transverse charge distribution as a
function of the impact parameter is simply the two-dimensional Fourier
transform of the Dirac form factor $F_1(Q^2)$.  This provides a
model-independent extraction of the transverse spatial distribution of charge
for a nucleon in the IMF, which is more closely connected to the quark parton
momentum distributions and GPDs.  While the proton's transverse charge
distribution was consistent with expectations, the neutron yielded a small,
negative region of charge in the very center.  This novel feature was
not consistent with expectations based on simply treating the IMF distribution
as the transverse spatial distribution in the rest frame.

\begin{figure}[hbt]
\begin{center}
\epsfig{file=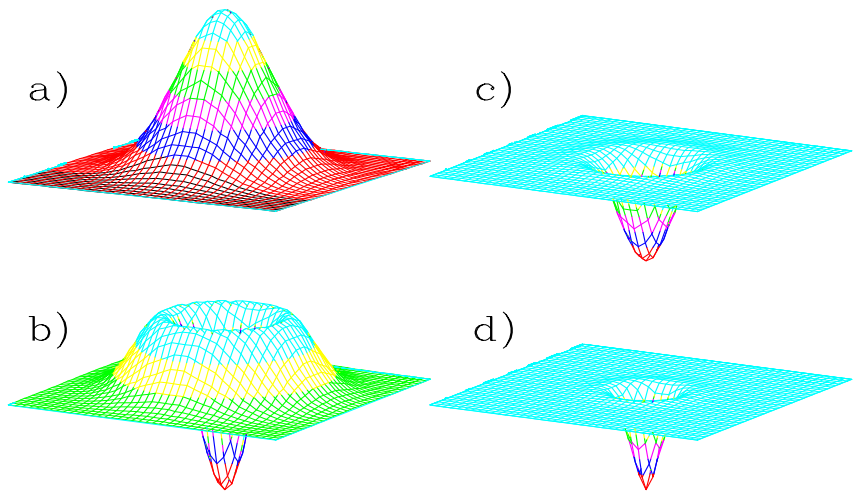,height=0.7\linewidth,width=0.4\linewidth,angle=90}
\end{center}
\caption{Contributions to the infinite momentum frame transverse charge
density of the neutron coming from low (a) to high (d) momentum
quarks~\cite{miller08b}.}
\label{fig:neutron_density}
\end{figure}

The Drell-Yan-West relation links the $x$-dependence of a parton distribution
function $q(x)$ at large $x$ to the $Q^2$-dependence of a form factor $F(Q^2)$:
\begin{equation} 
q(x) \propto (1-x)^{(\nu - 1)} ~ \longleftrightarrow ~  F(Q^2) \propto Q^{-\nu}.
\label{eq:dyw}
\end{equation}
The preliminary $\gen$ data at large \q2 show a slower fall-off with $Q^2$ than
the $\gep$ data, indicating a dominance of $d$ quarks over $u$ quarks at large
$x$. This observation, first suggested by Kroll~\cite{kroll07}, is in
agreement with the phenomenological modeling of the parton distribution
functions, such as CTEQ6M~\cite{pumplin02}. Using models of neutron GPDs, the
correlation between the transverse spatial distributions of the nucleon
in the IMF and
the quark momenta was examined, and it was shown that this negative central
charge in the neutron was the result of this strong $d$-quark dominance at
high $x$~\cite{miller08b}.  In the IMF, the center of momentum of the nucleon
is the longitudinal momentum weighted average of the transverse quark
positions.  In the limit where one quark has $x \to 1$, the position of that
single quark provides the dominant contribution in defining the transverse
center of the nucleon, and thus contributions from large $x$ become localized
near the center of the nucleon. Figure~\ref{fig:neutron_density} shows the
charge density of the neutron coming from different $x$ regions. At low $x$,
the density of up and down quarks is similar, and there is a net positive
charge with a broad spatial distribution.  At large $x$, the density of down
quarks is more than twice that of up quarks, so the net charge is negative and
the distribution becomes more localized, yielding the small negative core in
the neutron.

Such visualizations of the nucleon charge distributions on the light front
have been expanded to the case of nucleons polarized transversely to the light
front~\cite{carlson08}. Both the (proton and neutron) unpolarized
distributions are shifted along the polarization axis due to the nucleon's
large anomalous magnetic moment, which results in an induced electric dipole
moment perpendicular to the polarization axis.

The new form-factor data provide information on much more than just the
spatial structure of the nucleon. The new high-$Q^2$ measurements, combined
with the new, complete set of low-$Q^2$ form-factor data, provide powerful
constraints on models of nucleon structure.  Figure~\ref{fig:scaling} shows a
small selection of calculations and it is clear that there is a significant
scatter even between modern calculations.  Prior to the Jefferson Lab program, only
$\gmp$ was known with high precision over a large $Q^2$-range; the other form-factor 
extractions were limited in both precision and kinematic coverage. 
Many theoretical approaches were used to model nucleon structure and with
parameters adjusted to reproduce $\gmp$, there were only limited constraints
from the other form factors.  The improved measurements of $\gep$ and $\gmn$
provided real challenges for calculations that had, until this time, been
minimally constrained.  High-precision measurements of $\gen$ have
been more difficult to incorporate in evaluating models of nucleon structure.
Pion-cloud contributions are neglected in many calculations, but play an
important role in the behavior of $\gen$ at low $Q^2$.  The newest $\gen$
measurement extends the data to 3.4~GeV$^2$, providing a complete set of
form-factor extractions in the $Q^2$-region where pion-cloud contributions
should be small and the form factor is expected to be sensitive to the quark
core of the neutron.

Early experiments on the proton focused on the high-$Q^2$ region, but precision
measurements at low $Q^2$ also became an area of renewed interest. Had the up
and down quark distributions been identical, the neutron would have been
neutral everywhere yielding $\gen=0$ at all $Q^2$-values.  The non-zero values
of $\gen$ at low $Q^2$ demonstrate that there is a net positive charge
in the core of the neutron and a negative charge distribution on the outside,
as expected from virtual fluctuations of the neutron into a proton and a
negative pion.  An intriguing analysis~\cite{friedrich03} suggested the
possibility that similar contributions may be present in all of the form
factors.  For the proton, the pion cloud is a small contribution on top of the
quark core, and precision data are necessary to test models of pion
contributions for all four form factors.  New measurements~\cite{crawford07,
ron07} at low $Q^2$, shown in Fig.~\ref{fig:gepgmp} showed hints of structure
in the ratio $\mugegmp$, motivating experiment E08-007, which will cover this
$Q^2$-range with significantly higher precision.

Having a precise and complete data set at low $Q^2$ also provides the
opportunity to study the contributions of the different quark flavors. 
Neglecting strange and heavier quarks, the proton and neutron form factors
consist of contributions from the up and down quarks. Since many nucleon
properties appear to have significant contributions from strangeness, neglecting
strange quarks may not be well justified, and additional information is required
to constrain the impact of strangeness on the nucleon form factors. 
Measurements of parity-violating electron scattering can provide information
on the coupling of a Z-boson to the proton. 
This provides a complete set of observables that allows for a separation of
the up, down, and strange quark contributions to the nucleon form factors, as
discussed in detail in the following article~\cite{paschke11}.  Data on both proton
and neutron are also needed to extract the isovector form factors (proton minus
neutron).  This is the combination that can be most reliably extracted from 
lattice QCD calculations, where taking the difference between proton and
neutron removes the influence of disconnected diagrams which are difficult to
calculate.

Finally, data at even lower $Q^2$-values, below 0.1~GeV$^2$, can also have an
important impact. Extraction of the proton charge radius depends on having
reliable measurements of the form factors at extremely low $Q^2$, and what we
have learned about two-photon exchange corrections has led to an updated
extraction of the proton charge radius~\cite{rosenfelder00,blunden05b}. Ultra-high
precision atomic physics measurements, e.g. hyperfine splitting in hydrogen
and muonic hydrogen, are extremely sensitive to proton structure corrections,
including terms directly calculated from the low-$Q^2$ proton form
factors~\cite{carlson08}.  Future experiments (Table~\ref{tab:experiments})
will extend proton form factor extractions down below 0.02~GeV$^2$ with significantly
improved precision over older Rosenbluth experiments.

\section{Future Plans and Outlook}\label{sec:summary}

State of the art polarized electron beams, coupled with high figure-of-merit
polarized targets and recoil polarimeters, have enabled a program of
measurements at Jefferson Lab that has dramatically modified our picture of
nucleon form factors. Individual experiments, in particular the surprising
results of the proton electric form-factor measurements, have led to a
reexamination of long held pictures of nucleon structure.  Taken together,
these experiments have provided data of dramatically improved quality for most
form factors over large kinematic ranges, and have led to a
resurgence in efforts to evaluate nucleon models against a complete set
of form factor measurements.  With neutron measurements beginning to cover
a significant portion of the $Q^2$-range for which precise proton data already
exist, we can begin to extract model-independent information about the
difference between the up and down quark distributions, and the high-$Q^2$
extractions of $\gen$ show that there is a noticeable difference between
the up and down quark spatial distributions in the neutron~\cite{riordan10}.

These new data, as well as the techniques that made these measurements
possible, will continue to have impact on other experimental investigations.
Precise knowledge of the electromagnetic form factors is necessary to probe
the strange quark contributions to the nucleon using parity-violating
lepton scattering~\cite{paschke11}.  It is also important input to
high-precision nuclear structure measurements that utilize quasi-elastic
scattering from the nucleus, and measurements of hyperfine splitting in
hydrogen.  In addition, the measurement of nucleon form factors utilizing
polarization measurements allows for much more sensitive investigations into
the effect of the nuclear medium on the nucleon form factors.  The comparison
of free to bound proton form factors, using polarization transfer in
quasi-elastic scattering from $^4$He~\cite{strauch02,brooks11}, yields smaller
corrections than similar measurements that relied on Rosenbluth separations,
although the impact of these corrections on the interpretation is a topic of
great interest~\cite{schiavilla05}.

Soon, the last of the measurements from Table~\ref{tab:experiments} will be
completed as the 6 GeV program is brought to a close. The increased electron
energies available after the 12 GeV upgrade, coupled with further improvements
in the experimental equipment, will enable a dramatic extension of the
program presented here, doubling or tripling the $Q^2$ range of most of these
measurements.  This will provide valuable constraints on generalized parton
distributions, an important focus of the upgrade, at very high $Q^2$ values.
In addition to the benefits gained by extending the $Q^2$-range for the
individual form factors, this will also provide a complete set of form factor
measurements at large $Q^2$, where the pion-cloud contributions are expected
to be small and the measurements can be directly compared to calculations of
the quark core of the nucleon. This will make evaluation of nucleon models
more reliable, as pion-cloud contributions are typically difficult to include
in a self-consistent fashion.

\ack
This work was supported by the Department of Energy, Office of Nuclear
Physics, contract nos. DE-AC05-84ER40150 and DE-AC02-06CH11357 and by the
National Science Foundation, grant PHY-0753777. The authors thank those of
their colleagues who assisted in the preparation of this overview and
apologies for omissions made necessary by the constraints of length and time.

\section*{References}
\bibliographystyle{test6}
\bibliography{jlab_form_factors}

\begin{thebibliography}{10}
\providecommand{\url}[1]{\texttt{#1}}
\providecommand{\urlprefix}{URL }
\providecommand{\eprint}[2][]{\url{#2}}

\bibitem{hydewright04}
C.~E. Hyde-Wright and K.~de~Jager, Ann. Rev. Nucl. Part. Sci. \textbf{54}, 217
  (2004).

\bibitem{perdrisat07}
C.~F. Perdrisat, V.~Punjabi and M.~Vanderhaeghen, Prog. Part. Nucl. Phys.
  \textbf{59}, 694 (2007).

\bibitem{arrington07a}
J.~Arrington, C.~D. Roberts and J.~M. Zanotti, J. Phys. \textbf{G34}, S23
  (2007).

\bibitem{platchkov90}
S.~Platchkov \emph{et~al.}, Nucl. Phys. \textbf{A510}, 740 (1990).

\bibitem{akhiezer58}
A.~I. Akhiezer, L.~N. Rozentsweig and I.~M. Shmuskevich, Sov. Phys. JETP
  \textbf{6}, 588 (1958).

\bibitem{akhiezer68}
A.~I. Akhiezer and M.~P. Rekalo, Sov. Phys. Dokl. \textbf{13}, 572 (1968).

\bibitem{dombey69}
N.~Dombey, Rev. Mod. Phys. \textbf{41}, 236 (1969).

\bibitem{arnold81}
R.~G. Arnold, C.~E. Carlson and F.~Gross, Phys. Rev. C \textbf{23}, 363 (1981).

\bibitem{punjabi05}
V.~Punjabi \emph{et~al.}, Phys. Rev. C \textbf{71}, 055202 (2005),
  erratum-ibid. C71:069902, 2005.

\bibitem{jones00}
M.~K. Jones \emph{et~al.}, Phys. Rev. Lett. \textbf{84}, 1398 (2000).

\bibitem{gayou02}
O.~Gayou \emph{et~al.}, Phys. Rev. Lett. \textbf{88}, 092301 (2002).

\bibitem{puckett10}
A.~J.~R. Puckett \emph{et~al.}, Phys. Rev. Lett. \textbf{104}, 242301 (2010).

\bibitem{arrington04a}
J.~Arrington, Phys. Rev. C \textbf{69}, 022201(R) (2004).

\bibitem{qattan05}
I.~A. Qattan \emph{et~al.}, Phys. Rev. Lett. \textbf{94}, 142301 (2005).

\bibitem{crawford07}
C.~B. Crawford \emph{et~al.}, Phys. Rev. Lett. \textbf{98}, 052301 (2007).

\bibitem{paolone10}
M.~Paolone, S.~Malace, S.~Strauch, I.~Albayrak, J.~Arrington \emph{et~al.},
  Phys.Rev.Lett. \textbf{105}, 072001 (2010).

\bibitem{zhan11}
X.~Zhan \emph{et~al.}, \textit{arXiv:nucl-ex/1102.0318}  (2011).

\bibitem{ron11}
G.~Ron \emph{et~al.}, \textit{to be submitted to Phys. Rev. C}  (2011).

\bibitem{kelly04}
J.~J. Kelly, Phys. Rev. C \textbf{70}, 068202 (2004).

\bibitem{arrington07c}
J.~Arrington, W.~Melnitchouk and J.~A. Tjon, Phys. Rev. \textbf{C76}, 035205
  (2007).

\bibitem{arrington03a}
J.~Arrington, Phys. Rev. C \textbf{68}, 034325 (2003).

\bibitem{christy04}
M.~E. Christy \emph{et~al.}, Phys. Rev. C \textbf{70}, 015206 (2004).

\bibitem{afanasev01a}
A.~Afanasev, I.~Akushevich and N.~Merenkov, Phys. Rev. \textbf{D64}, 113009
  (2001).

\bibitem{maximon00}
L.~C. Maximon and J.~A. Tjon, Phys. Rev. C \textbf{62}, 054320 (2000).

\bibitem{guichon03}
P.~A.~M. Guichon and M.~Vanderhaeghen, Phys. Rev. Lett. \textbf{91}, 142303
  (2003).

\bibitem{friedrich03}
J.~Friedrich and T.~Walcher, Eur. Phys. J. \textbf{A17}, 607 (2003).

\bibitem{ron07}
G.~Ron \emph{et~al.}, Phys. Rev. Lett. \textbf{99}, 202002 (2007).

\bibitem{bernauer10}
J.~Bernauer \emph{et~al.}, Phys.Rev.Lett. \textbf{105}, 242001 (2010).

\bibitem{blunden03}
P.~G. Blunden, W.~Melnitchouk and J.~A. Tjon, Phys. Rev. Lett. \textbf{91},
  142304 (2003).

\bibitem{arrington04b}
J.~Arrington, Phys. Rev. C \textbf{69}, 032201(R) (2004).

\bibitem{arrington04d}
J.~Arrington, Phys. Rev. C \textbf{71}, 015202 (2005).

\bibitem{meziane10}
M.~Meziane \emph{et~al.}, \textit{arXiv:nucl-ex/1012.0339}  (2010).

\bibitem{guttmann10}
J.~Guttmann, N.~Kivel, M.~Meziane and M.~Vanderhaeghen,
  \textit{arXiv:hep-ph/1012.0564}  (2010).

\bibitem{borisyuk10}
D.~Borisyuk and A.~Kobushkin, \textit{arXiv:hep-ph/1012.3746}  (2010).

\bibitem{carlson07}
C.~E. Carlson and M.~Vanderhaeghen, Ann. Rev. Nucl. Part. Sci. \textbf{57}, 171
  (2007).

\bibitem{blunden05a}
P.~G. Blunden, W.~Melnitchouk and J.~A. Tjon, Phys. Rev. C \textbf{72}, 034612
  (2005).

\bibitem{kondratyuk07}
S.~Kondratyuk and P.~G. Blunden, Phys. Rev. C \textbf{75}, 038201 (2007).

\bibitem{chen04}
Y.~C. Chen, A.~Afanasev, S.~J. Brodsky, C.~E. Carlson and M.~Vanderhaeghen,
  Phys. Rev. Lett. \textbf{93}, 122301 (2004).

\bibitem{afanasev05a}
A.~V. Afanasev, S.~J. Brodsky, C.~E. Carlson, Y.-C. Chen and M.~Vanderhaeghen,
  Phys. Rev. D \textbf{72}, 013008 (2005).

\bibitem{afanasev02a}
A.~V. Afanasev, I.~Akushevich and N.~P. Merenkov, Phys. Rev. \textbf{D65},
  013006 (2002).

\bibitem{bystritskiy07}
Y.~M. Bystritskiy, E.~A. Kuraev and E.~Tomasi-Gustafsson, Phys. Rev.
  \textbf{C75}, 015207 (2007).

\bibitem{afanasev07}
A.~V. Afanasev, \textit{Proceedings of the International Workshop on Exclusive
  Reactions at High Momentum Transfer, 21-24 May 2007, Jefferson Lab.
  arXiv:hep-ph/0711.3065}  (2008).

\bibitem{mo69}
L.~W. Mo and Y.-S. Tsai, Rev. Mod. Phys. \textbf{41}, 205 (1969).

\bibitem{tvaskis06}
V.~Tvaskis \emph{et~al.}, Phys. Rev. C \textbf{73}, 025206 (2006).

\bibitem{pentchev08}
L.~Pentchev, AIP Conf. Proc. \textbf{1056}, 357 (2008).

\bibitem{vepp_proposal}
J.~Arrington, D.~M. Nikolenko \emph{et~al.}, Proposal for positron measurement
  at VEPP-3, \eprint{nucl-ex/0408020}.

\bibitem{arrington09b}
J.~Arrington, AIP Conf. Proc. \textbf{1160}, 13 (2009), \eprint{0905.0713}.

\bibitem{anderson07}
B.~Anderson \emph{et~al.}, Phys. Rev. C \textbf{75}, 034003 (2007).

\bibitem{anklin98}
H.~Anklin \emph{et~al.}, Phys. Lett. \textbf{B428}, 248 (1998).

\bibitem{kubon02}
G.~Kubon \emph{et~al.}, Phys. Lett. \textbf{B524}, 26 (2002).

\bibitem{lachniet08}
J.~Lachniet \emph{et~al.}, Phys. Rev. Lett. \textbf{102}, 192001 (2009).

\bibitem{jeschonnek00}
S.~Jeschonnek and J.~W. Van~Orden, Phys. Rev. C \textbf{62}, 044613 (2000).

\bibitem{xu00}
W.~Xu \emph{et~al.}, Phys. Rev. Lett. \textbf{85}, 2900 (2000).

\bibitem{xu03}
W.~Xu \emph{et~al.}, Phys. Rev. C \textbf{67}, 012201 (2003).

\bibitem{alcorn04}
J.~Alcorn \emph{et~al.}, Nucl. Inst. \& Meth. \textbf{A522}, 294 (2004).

\bibitem{golak01}
J.~Golak, G.~Ziemer, H.~Kamada, H.~Witala and W.~Gloeckle, Phys. Rev. C
  \textbf{63}, 034006 (2001).

\bibitem{madey03}
R.~Madey \emph{et~al.}, Phys. Rev. Lett. \textbf{91}, 122002 (2003).

\bibitem{plaster06}
B.~Plaster \emph{et~al.}, Phys. Rev. \textbf{C73}, 025205 (2006).

\bibitem{warren04}
G.~Warren \emph{et~al.}, Phys. Rev. Lett. \textbf{92}, 042301 (2004).

\bibitem{crabb95}
D.~G. Crabb and D.~B. Day, Nucl. Instrum. Meth. \textbf{A356}, 9 (1995).

\bibitem{averett99}
T.~D. Averett \emph{et~al.}, Nucl. Instrum. Meth. \textbf{A427}, 440 (1999).

\bibitem{schiavilla01}
R.~Schiavilla and I.~Sick, Phys. Rev. C \textbf{64}, 041002(R) (2001).

\bibitem{geis08}
E.~Geis \emph{et~al.}, Phys. Rev. Lett. \textbf{101}, 042501 (2008).

\bibitem{riordan10}
S.~Riordan \emph{et~al.}, Phys. Rev. Lett. \textbf{105}, 262302 (2010).

\bibitem{belitsky03}
A.~V. Belitsky, X.-d. Ji and F.~Yuan, Phys. Rev. Lett. \textbf{91}, 092003
  (2003).

\bibitem{miller02b}
G.~A. Miller, Phys. Rev. C \textbf{66}, 032201 (2002).

\bibitem{cloet08}
I.~C. Cloet, G.~Eichmann, B.~El-Bennich, T.~Klahn and C.~D. Roberts, Few Body
  Syst. \textbf{46}, 1 (2009).

\bibitem{bijker04}
R.~Bijker and F.~Iachello, Phys. Rev. C \textbf{69}, 068201 (2004).

\bibitem{lomon02}
E.~L. Lomon, Phys. Rev. C \textbf{66}, 045501 (2002).

\bibitem{lomon06}
E.~L. Lomon, \textit{arXiv:nucl-th/0609020}  (2006), \eprint{nucl-th/0609020}.

\bibitem{hammer04b}
H.~W. Hammer and U.-G. Meissner, Eur. Phys. J. \textbf{A20}, 469 (2004).

\bibitem{kelly02}
J.~J. Kelly, Phys. Rev. C \textbf{66}, 065203 (2002).

\bibitem{miller03}
G.~A. Miller, Phys. Rev. C \textbf{68}, 022201 (2003).

\bibitem{belitsky04}
A.~V. Belitsky, X.-d. Ji and F.~Yuan, Phys. Rev. \textbf{D69}, 074014 (2004).

\bibitem{venkat11}
S.~Venkat, J.~Arrington, G.~A. Miller and X.~Zhan, Phys. Rev. C \textbf{83},
  015203 (2011).

\bibitem{hyde11}
C.~E. Hyde, M.~Guidal and A.~V. Radyushkin, {\textit{Deep Virtual Exclusive
  Processes and Generalized Parton Distributions}, this volume, 2011}.

\bibitem{miller07}
G.~A. Miller, Phys. Rev. Lett. \textbf{99}, 112001 (2007).

\bibitem{kroll07}
P.~Kroll, \textit{Proceedings of the International Workshop on Exclusive
  Reactions at High Momentum Transfer, 21-24 May 2007, Jefferson Lab.
  arXiv:hep-ph/0710.2771}  (2008).

\bibitem{pumplin02}
J.~Pumplin \emph{et~al.}, JHEP \textbf{07}, 012 (2002).

\bibitem{miller08b}
G.~A. Miller and J.~Arrington, Phys. Rev. \textbf{C78}, 032201 (2008).

\bibitem{carlson08}
C.~E. Carlson, V.~Nazaryan and K.~Griffioen, arXiv:physics.atom-ph/0805.2603
  (2008), \eprint{0805.2603}.

\bibitem{paschke11}
K.~Paschke, A.~Thomas, R.~Michaels and D.~Armstrong, {\textit{Strange Vector
  Form-Factors from Parity-Violating Electron Scattering}, this volume, 2011}.

\bibitem{rosenfelder00}
R.~Rosenfelder, Phys. Lett. B \textbf{479}, 381 (2000).

\bibitem{blunden05b}
P.~G. Blunden and I.~Sick, Phys. Rev. C \textbf{72}, 057601 (2005).

\bibitem{strauch02}
S.~Strauch \emph{et~al.}, Phys. Rev. Lett. \textbf{91}, 052301 (2003).

\bibitem{brooks11}
W.~K. Brooks, S.~Strauch and K.~Tsushima, {\textit{Medium Modifications of
  Hadron Properties and Partonic Processes}, this volume, 2011}.

\bibitem{schiavilla05}
R.~Schiavilla, O.~Benhar, A.~Kievsky, L.~E. Marcucci and M.~Viviani, Phys. Rev.
  Lett. \textbf{94}, 072303 (2005).

\end{thebibliography}

\end{document}